\documentclass[journal]{IEEEtran}
\usepackage[linesnumbered,ruled,lined]{algorithm2e}%-- 算法流程包
\usepackage{graphicx}
\usepackage{latexsym}
\usepackage{amsfonts}
\usepackage{subfigure}
\usepackage{dsfont}
\usepackage{epstopdf}
\usepackage{threeparttable}
\usepackage{multirow}
\usepackage{enumerate}
\usepackage{epsfig}
\usepackage[justification=centering]{caption}
\usepackage{bm}
\usepackage{cite}
\usepackage{textcomp}
\usepackage{booktabs}
\usepackage{tabu}
\usepackage{amsmath}
\usepackage{subfig}
\usepackage{bm}%--\gama加粗包
\usepackage{harpoon}

\usepackage{enumerate}
\usepackage{enumitem}

\makeatletter

\newcommand{\Rmnum}[1]{\expandafter\@slowromancap\romannumeral #1@}
\makeatother

\usepackage{amsmath}

\usepackage{graphicx}
\usepackage{epstopdf}
\usepackage{subfigure}

\usepackage[justification=centering]{caption}

\usepackage{setspace}

\usepackage{enumerate}

\usepackage{amsfonts}

\usepackage{booktabs}

\usepackage{color}

\usepackage{cite}

\begin{document}

% ====letter title===========
\title{Cascaded ELM-based Joint Frame Synchronization and Channel Estimation over Rician Fading Channel with Hardware Imperfections}

\author{Chaojin~Qing,~\IEEEmembership{Member,~IEEE,}
        Chuangui~Rao,
        Shuhai Tang,
        Na Yang
        and Jiafan Wang %<-this  stops a space

\thanks{This work is supported in part by the Sichuan Science and Technology Program (Grant No. 2021JDRC0003, 2021YFG0064), the Major Special Funds of Science and Technology of Sichuan Science and Technology Plan Project (Grant No. 19ZDZX0016 /2019YFG0395), and the Demonstration Project of Chengdu Major Science and Technology Application (Grant No. 2020-YF09- 00048-SN).}

\thanks{Chaojin Qing, Chuangui Rao, Shuhai Tang, Na Yang, and Jiafan Wang are with the School of Electrical Engineering and Electronic Information, Xihua University, Chengdu, 610039, China (E-mail: qingchj@mail.xhu.edu.cn, raochuangui5621@163.com, tangshh@stu.xhu.edu.cn, yangna6717@163.com, and jifanw@gmail.com).}

}

\markboth{XXXX}%
{Qing \MakeLowercase{\textit{et al.}}: ELM-BASED FRAME SYNCHRONIZATION IN BURST-MODE COMMUNICATION SYSTEMS WITH NONLINEAR DISTORTION}

\maketitle

%===========摘要==============
\begin{abstract}
Due to the interdependency of frame synchronization (FS) and channel estimation (CE), joint FS and CE (JFSCE) schemes are proposed to enhance their functionalities and therefore boost the overall performance of wireless communication systems. Although traditional JFSCE schemes alleviate the influence between FS and CE, they show deficiencies in dealing with hardware imperfection (HI) and deterministic line-of-sight (LOS) path. To tackle this challenge, we proposed a cascaded ELM-based JFSCE to alleviate the influence of HI in the scenario of the Rician fading channel. Specifically, the conventional JFSCE method is first employed to extract the initial features, and thus forms the non-Neural Network (NN) solutions for FS and CE, respectively. Then, the ELM-based networks, named FS-NET and CE-NET, are cascaded to capture the NN solutions of FS and CE. Simulation and analysis results show that, compared with the conventional JFSCE methods, the proposed cascaded ELM-based JFSCE significantly reduces the error probability of FS and the normalized mean square error (NMSE) of CE, even against the impacts of parameter variations.
\end{abstract}

%========= 关键字============

\IEEEpeerreviewmaketitle

%=========第一章introduction （空一行表示分段）============
\section{Introduction}
\IEEEPARstart{F}{rame} synchronization (FS) and channel estimation (CE) are crucial tasks in wireless communication systems due to their roles in successful data reception \cite{FSCE1, FSCE2}. Usually, FS and CE are interdependent \cite{flat_fading,ref_ML, ref_OMP}, and thus the joint FS and CE (JFSCE) are inspired. In \cite{flat_fading}, a JFSCE is proposed for relay networks in flat fading channels, and achieves accurate estimation for CE and timing offset. In \cite{ref_ML}, a computationally efficient continuous-mode algorithm for JFSCE is proposed in frequency-selective channels, in which the FS's error probability and CE's normalized mean square error (NMSE) are reduced. In \cite{ref_OMP}, the sparsity reconstruction-based JFSCE is proposed to cope with the significant performance degradation of correlation-based FS. From \cite{flat_fading,ref_ML, ref_OMP}, both the FS's error probability and CE's NMSE are reduced, compared with independent FS and CE.

Nevertheless, the JFSCE methods in \cite{flat_fading,ref_ML, ref_OMP} are still facing severe challenges in real wireless communication systems due to hardware imperfections (HI). The HI usually comes from various components, such as digital to analog converter (DAC), high power amplifier (HPA), mixers, and filters, etc., \cite{HI_r1, HI_r2}. Limited battery energy and insufficient computational resources \cite{low_cost, energy, compute} lead to the HI and cause serious nonlinear distortion to the received signals, resulting in perennial impact on the system performance \cite{c9, c10, c11}. For JFSCE, however, HI has not been considered in the existing methods, e.g., \cite{flat_fading,ref_ML, ref_OMP}, making it difficult for the method to be practical. Thus, this issue has to be addressed in JFSCE.

To address the issue arising from HI, machine learning (ML), especially deep neural network (DNN) and extreme learning machine (ELM), presents its significant superiority \cite{MLtoHI2, MLtoHI3, arXiv200709248L, OFDM_time_synchron, ELM_fs1,ELM_fs2}. In \cite{MLtoHI2}, the neural network (NN) technique is employed for CE to alleviate the symbol misidentification caused by HI. Then, a deep residual learning-based joint pilot and CE method is proposed in \cite{MLtoHI3}, which mitigates the influences of pilot contamination but skips the JFSCE method. For other literature \cite{arXiv200709248L, OFDM_time_synchron, ELM_fs1,ELM_fs2}, the ELM-based synchronization is investigated. In \cite{arXiv200709248L}, an ELM-based fine timing synchronization (TS) is analyzed based on the existing coarse synchronization and CE, while the HI is with little concern. By contrast, an ELM-based TS proposed in  \cite{OFDM_time_synchron} factors in the HI by exploiting the features of learning labels and thus achieves the improvement in the error probability of TS. In \cite{ELM_fs1}, an ELM-based FS in burst-mode communication systems with HI is investigated. Relative to the classic cross-correlation-based FS \cite{YW_ref1,YW_ref2}, this FS \cite{ELM_fs1} presents a reduction in the error probability of FS and possesses good robustness against the varying parameters of HI. The ELM-based FS in \cite{ELM_fs1} is extended to the superimposed FS in \cite{ELM_fs2} to improve the spectral efficiency during the FS phase. From \cite{arXiv200709248L, OFDM_time_synchron, ELM_fs1,ELM_fs2}, the ELM network is introduced into synchronization due to the following considerations. 1) As a feed-forward neural network, ELM does not need to adjust the network weight based on the gradient back propagation, speeding its training. 2) ELM possesses low computational complexity due to its lightweight network structure with a single hidden layer, which facilitates its practical applications in the scenario of limited computing resources. 3) In many existing works, ELM embodies excellent generalization, e.g., \cite{huang2004extreme, huang2006extreme}. However, FS is affected by CE and vice versa. In \cite{arXiv200709248L, OFDM_time_synchron, ELM_fs1,ELM_fs2}, the influences between TS/FS and CE are not considered, severely degrading system performance. Although the ELM networks are introduced into FS, the ELM-based JFSCE has not been investigated. Therefore, the JFSCE and HI should be considered in the real wireless communication systems, motivating us to tackle this issue in this paper.

In addition, in \cite{flat_fading,ref_ML,ref_OMP,ELM_fs1,ELM_fs2,OFDM_time_synchron,arXiv200709248L}, the flat/frequency-selective fading channels are considered, omitting the line-of-sight (LOS) component. Yet still, in practical and more widely used scenarios, e.g., indoor radio channels involving deterministic LOS path component\cite{indoors1,distance_shorter}, air-to-ground channels with less shadow fading \cite{A2G}, etc, the possibility of existing LOS paths is significantly increased. Especially in the fifth-generation and sixth-generation communication systems, to satisfy the demands of users' traffic and quality-of-service, the distance between the base station (or access point) and the user equipment is becoming closer than that of the past mobile communication systems. This leads to the LOS path more deterministic \cite{distance_shorter,millimeterWave,rice_discrip}, making the Rician fading would be considered as an important scenario in real wireless communication systems.

Considering the scenarios of Rician fading channels with HI, a cascaded ELM-based JFSCE method is proposed in this paper. The main contributions are summarized as follows:

\begin{enumerate}
  \item We develop an ELM-based JFSCE method with HI. In most existing works, either HI is not considered in JFSCE, or the joint processing between FS and CE is not employed. Thus, on the one hand, the methods of JFSCE, e.g., \cite{flat_fading,ref_ML, ref_OMP}, suffer from severe performance degradation due to the lack of consideration for the HI in the real wireless systems. On the other hand, without the joint processing, ELM-based FSs in \cite{arXiv200709248L, OFDM_time_synchron, ELM_fs1,ELM_fs2} are significantly affected by CE. Both the JFSCE and HI are considered in this paper, and the ELM networks are developed to address these issues.

  \item We exploit a strategy to fuse the neural network (NN) and non-NN for the ELM-based JFSCE. Our key insight is to view the HI in JFSCE as a nonlinear problem and solve this problem by fusing the non-NN and NN-based modes. Specifically, the non-NN based JFSCE method in \cite{ref_OMP} is first employed to capture the initial features of FS and CE, respectively. Then, the ELM-based networks, named FS-NET and CE-NET, are respectively constructed to refine the FS and CE. Continuing with the non-NN receiver solutions, the FS-NET and CE-NET reduce the FS's error probability and CE's NMSE.

  \item We consider the LOS component of wireless channels, and thus the Rician fading channel is employed to investigate the ELM-based JFSCE, facilitating its practical applications. In \cite{flat_fading,ref_ML, ref_OMP}, the JFSCE methods were investigated in flat/requency-selective fading channels, omitting the LOS components of the wireless channel. To this end, the proposed scheme can not only work in the flat/frequency-selective fading channels but also in the Rician fading channels.

  \item We form a paradigm by using the cascaded ELM networks to solve the multi-task problems. As the practical demonstration in this paper, the ELM-based FS-NET and CE-NET are cascaded to perform the FS and CE, respectively. This paradigm presents many benefits, such as accelerating network training, avoiding complex parameter tuning, etc. Therefore, this paradigm is suitable for solving multi-task problems. In this work, the FS, CE, and HI's suppression are benefited from this paradigm.

\end{enumerate}

The remainder of this paper is structured as follows. In Section II, the system model is described. The ELM-based JFSCE algorithm is given in Section III, followed by the numerical simulation and analysis in Section IV. Besides, the Section V concludes our work.

Notations: Bold lowercase and uppercase letters denote vectors and matrices, respectively; italicized letters denote variables; $(\cdot)^{T}$, $(\cdot)^{H}$, $(\cdot)^{-1}$ and ${\left( \cdot\right)^\dag }$ denote the transpose, conjugate transpose, matrix inversion, Moore-Penrose pseudoinverse, respectively; ${\bf{0}}_{N}$ is ${N} \times 1$ vector with ${N}$ zero elements; ${\bf{I}}_N$ represents $N$-dimensional identity matrix; ${{\bf{X}}}\left( {:,N} \right)$ represents the $N$-th column of matrix ${{\bf{X}}}$; $\left| \cdot \right|$ denotes the absolute value or the absolute value operation; ${{\left\| \cdot  \right\|}_{2}}$ is the Frobenius norm; $\sigma \left( \cdot \right)$ denotes the activation function such as sigmoid, hyperbolic tangent, rectified linear units \cite{act_function}.

%==========第二章System model===============
\section{System Model}
\begin{figure}[t]
    \centering
    \includegraphics[width=1.0\columnwidth]{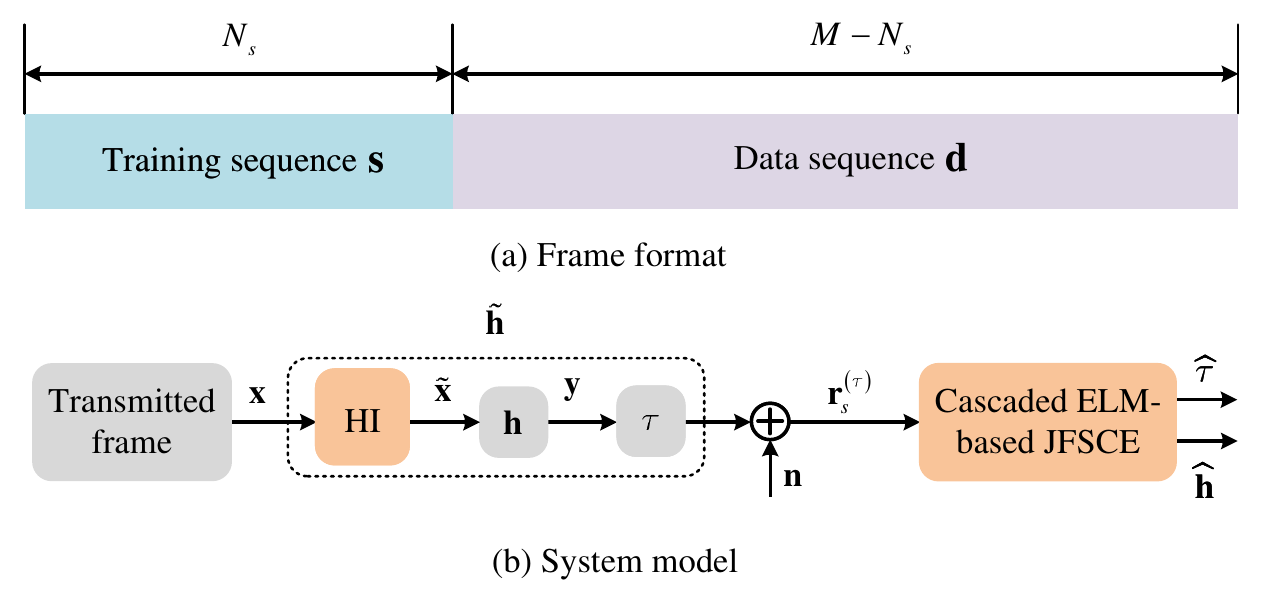}
    \caption{Frame format and proposed system model.}
    \label{system_model}
\end{figure}

%=============正文===================

The frame format and system model are given in Fig. \ref{system_model}(a) and Fig. \ref{system_model}(b), respectively. From Fig. \ref{system_model}(a), the $M$-length transmitted frame consists of $N_s$-length training sequence and data sequence with the length of $M-N_s$. In this paper, we denote the training sequence and data sequence as ${\bf{s=}}{\left[ {{s_0},{s_1}, \cdots ,{s_{{N_s} - 1}}} \right]^T}$ and ${\bf{d}} = {\left[ {{d_0},{d_1}, \cdots ,{d_{{M - {N_s}} - 1}}} \right]^T} $, respectively. From Fig. \ref{system_model}(b), the $M$-length transmitted frame encounters HI due to the nonlinear components or devices \cite{HI_r1, HI_r2,low_cost, energy, compute}. At the receiver, the received signal ${{\bf{r}}_s^{\left( \tau  \right)} \in \mathbb{C} {^{M \times 1}}}$ is given by
\begin{equation}
\label{eq_1}
{{\bf{r}}_s^{\left( \tau  \right)}} = {T^\tau }\left( {\bf{y}} \right) + {\bf{n}},
\end{equation}
where ${\tau}$ is the unknown FS offset to be estimated with ${0 \le \tau \le M - 1}$ \cite{ref_ML}; ${{T^\tau}\left( \cdot \right)}$ denotes the right cyclic shift operator, which is defined as ${T^\tau }\left( {\bf{y}} \right): = {\left[ {{y_{M - \tau }},{y_{M - \tau + 1}}, \cdots ,{y_0},{y_1}, \cdots } \right]^T}$; ${{\bf{y}} \in \mathbb{C}{^{M \times 1}}}$ represents the received signal without FS offset; ${\bf{n}} \in \mathbb{C}{^{M \times 1}}$ denotes the complex additive white Gaussian noise (AWGN) vector with zero mean and unit variance, i.e., ${\bf{n}} \sim \mathcal{CN}\left( {0,\textbf{I}_M} \right)$. In (\ref{eq_1}), $\bf{y}$ is expressed as
\begin{equation}
\label{eq_2}
{\bf{y}} = {\bf{\widetilde Xh}},
\end{equation}
where ${\bf{\widetilde X}} \in {\mathbb{C}} ^{M \times L}$ is the transmitted signal shift matrix experienced HI. In this paper, we employ HPA as an example to present the impact of HI. ${\bf{h}} \in \mathbb{C} {^{L \times 1}}$ represents the channel impulse response with ${L}$ resolvable wireless paths. For Rician fading scenarios, the entry in ${\bf{h}}$ is modeled as the independent identically distributed (i.i.d) complex Rician random variable, and ${\bf{h}}$ is given by \cite{rician_fading}
\begin{equation}
\label{eq_3}
{\bf{h}} = \sqrt {\frac{K}{{K + 1}}} {{\bf{h}}_{{\rm{LOS}}}} + \sqrt {\frac{1}{{K + 1}}} {{\bf{h}}_{{\rm{NLOS}}}},
\end{equation}
where ${K}$ is the Rician factor; ${{{\bf{h}}_{{\rm{LOS}}}} \in \mathbb{C} {^{L \times 1}}}$ and ${{{\bf{h}}_{{\rm{NLOS}}}} \in \mathbb{C} {^{L \times 1}}}$ denote the deterministic LOS and non LOS (NLOS) components, respectively. From \cite{distance_shorter}, the Rician factor ${K}$ denotes the power ratio of the deterministic component over the scattered component. The entries of ${{{\bf{h}}_{{\rm{NLOS}}}} \in \mathbb{C} {^{L \times 1}}}$ are the i.i.d complex Gaussian variables with zero-mean and unit variance, i.e., ${{\bf{h}}_{{\rm{NLOS}}}} \sim \mathcal{CN}\left( {0,{\bf{I}}}_L \right)$. For ease of expression, we express the transmitted signal shift matrix $\bf{\widetilde X}$ in (\ref{eq_2}) as
\begin{equation}
\label{eq_4}
{\bf{\widetilde X}} = \left[ {\begin{array}{*{20}{c}}
{\widetilde x_0^{\left( k \right)}}&{\widetilde x_{M - 1}^{\left( {k - 1} \right)}}& \cdots &{\widetilde x_{M - L + 1}^{\left( {k - 1} \right)}}\\
{\widetilde x_1^{\left( k \right)}}&{\widetilde x_0^{\left( k \right)}}& \cdots &{\widetilde x_{M - L + 2}^{\left( {k - 1} \right)}}\\
 \vdots & \vdots & \cdots & \vdots \\
{\widetilde x_{M - 1}^{\left( k \right)}}&{\widetilde x_{M - 2}^{\left( k \right)}}& \cdots &{\widetilde x_{M - L}^{\left( k \right)}}
\end{array}} \right],
\end{equation}
where the superscripts $k$ and $k-1$ denote the $k$-th and $\left( {k - 1} \right)$-th frames, respectively. By denoting the transmitted signals with imperfect and perfect hardware as ${\bf{\widetilde X}} = \left[ {{{{\bf{\widetilde x}}}_0},{{{\bf{\widetilde x}}}_1}, \cdots ,{{{\bf{\widetilde x}}}_{L - 1}}} \right]$ and ${\bf{ X}} = \left[ {{{\bf{x}}_0},{{\bf{x}}_1}, \cdots ,{{\bf{x}}_{L - 1}}} \right]$, respectively, then we have
\begin{equation}
\label{eq_5}
{{\bf{\widetilde x}}_i} = {f_{\bf{d}}}\left( {{{\bf{x}}_i}} \right),i = 0,1, \cdots ,L - 1,
\end{equation}
where ${{f_{\bf{d}}}\left( \cdot \right)}$ represents the mapping function from the hardware perfection to the HI. ${\bf{x}} = {\left[ {{x_0},{x_1}, \cdots ,{x_{M - 1}}} \right]^T}$ is the transmitted signal vector.

With the received signal ${{\bf{r}}_s^{\left( \tau \right)}}$, continuing with the non-NN receiver solutions, we propose a cascaded ELM-based JFSCE framework to achieve enhanced FS and CE. In the framework, the HI is viewed as a nonlinear problem, and this problem is solved by fusing the solutions of NN and non-NN. The proposed framework develops the cascaded ELM networks to solve the multi-task problems, which is easy to implement and thus facilitates its practical application.

%==========III 联合同步算法==========
\section{Cascaded ELM-based JFSCE}

In this section, the proposed cascaded ELM-based JFSCE is presented. We first briefly describe the non-NN-based JFSCEs in Section III-A to form the baseline methods, including the classic JFSCE in \cite{ref_ML} and the novel orthogonal matching pursuit (OMP)-based JFSCE in \cite{ref_OMP}. Then, in Section III-B, the proposed scheme is elaborated.

\subsection{Classic Non-NN-based JFSCE}
The methods of JFSCE in \cite{ref_ML} and \cite{ref_OMP} are employed as the demonstrations of non-NN-based JFSCEs in this subsection and elaborated as follows.

\textit{${\rm{(1)}}$ JFSCE in \cite{ref_ML}}: From \cite{ref_ML}, the offset estimation of FS is given by
\begin{equation}
\label{tao_est}
\widehat \tau = \mathop {\arg \max }\limits_{0 \le \tau \le M - 1} \left( {{\left( {{\bf{r}}_s^{\left( \tau \right)}} \right)^H} \left( {{{\bf{B}}_m} - {{\bf{I}}_M}} \right){\bf{r}}_s^{\left( \tau \right)}} \right).
\end{equation}
In (\ref{tao_est}), ${\bf{B}}_m$ is the projection matrix and is expressed as $\label{B_m} {{\bf{B}}_m} = {{\bf{S}}_m}{\bf{S}}_m^{ - 1}$ \cite{ref_ML}, where ${\bf{S}}_m \in \mathbb{C} {^{\left(N_s-L+1 \right) \times \left(N_s-L+1 \right)}}$ is the same as that of \cite{ref_ML} (see equation (4) in \cite{ref_ML}).
According to the estimated $\widehat \tau$ in (\ref{tao_est}), the estimation of ${\bf{h}}$ is given by
\begin{equation}
\label{h_hat1}
{\bf{\widehat h}}\left( {\widehat \tau } \right) = {{\bf{S}}_m^{ - 1}}{{\bf{R}}_s}\left( {:,\widehat \tau } \right),
\end{equation}
where ${{\bf{R}}_s} \in \mathbb{C} {^{M \times M}}$ is the received signal shift matrix according to ${{\bf{r}}_s^{\left( \tau \right)}}$. By denoting ${{\bf{r}}_s^{\left( \tau \right)}}$ as ${{\bf{r}}_s^{\left( \tau \right)}} = {\left[ {{r_0},{r_1}, \cdots, {r_{M - 1}}} \right]^T}$, ${\bf{R}}_s$ is constructed according to the column-wise shifting ${{\bf{r}}_s^{\left( \tau \right)}}$ with shift step being 1 and first column being ${{\bf{r}}_s^{\left( \tau \right)}}$. According to (\ref{tao_est}) and (\ref{h_hat1}), both the FS and CE can be estimated.
%---处理流程图---
\begin{figure*}[t]
    \centering
    \includegraphics[width=1.8\columnwidth]{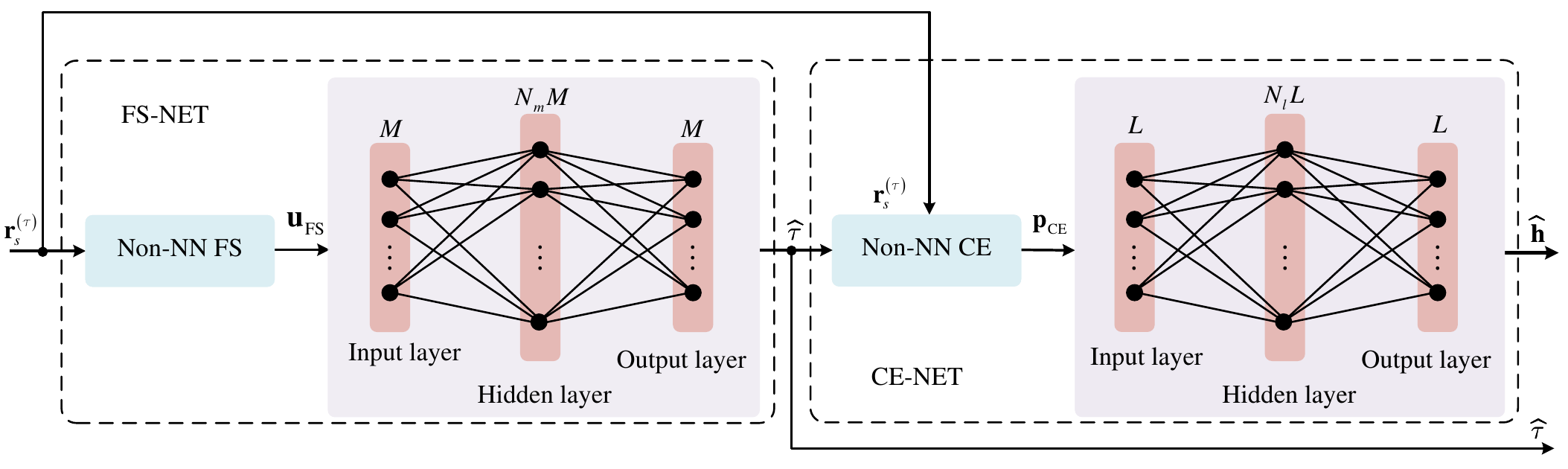}
    \caption{Cascaded ELM-based JFSCE.}
    \label{cascaded_network}
\end{figure*}

\textit{${\rm{(2)}}$ JFSCE in \cite{ref_OMP}}: In \cite{ref_OMP}, the OMP algorithm is exploited for JFSCE. Firstly, the classic cross-correlation is employed to form a synchronization metric (SM) ${\bf{u}} \in \mathbb{R} {^{M \times 1}}$, i.e.,
\begin{equation}
\label{sm}
{\bf{u}} = {{\bf{S}}^H}{{{\bf{r}}_s^{\left( \tau \right)}}},
\end{equation}
where ${\bf{S}} \in \mathbb{C} {^{M \times M}}$ is a Toeplitz matrix formed according to \cite{ELM_fs2} (see equation (7) in \cite{ELM_fs2}). Denoting ${\bf{u}} = {\left[ {{u_0},{u_1}, \cdots ,{u_{M - 1}}} \right]^T}$, then the estimation of $\tau$ is given by
\begin{equation}
\label{tao_est2}
\widehat \tau =\mathop{\arg \max} \limits_{0 \le i \le M - 1} {\left| {{u_{i}}} \right|^2}.
\end{equation}
With the estimated $\widehat \tau$, the CE is performed by employing the OMP algorithm, i.e.,
\begin{equation}
\label{h_hat2}
{{\bf{\widehat h}}} = {f_{{\rm{OMP}}}}\left( {{{{\bf{r}}_s^{\left( \tau \right)}}},{{\bf{X}}_\varphi }\left( {\widehat \tau } \right)} \right),
\end{equation}
where ${{\bf{X}}_\varphi }\left( {\widehat \tau } \right) \in \mathbb{C} {^{M \times L}}$ denotes the measurement matrix of the OMP algorithm.
The JFSCE in \cite{ref_OMP} employs classic cross-correlation to achieve the start point of FS and thus estimates the channel impulse response with the OMP algorithm.

In this paper, JFSCE methods in \cite{ref_ML} and \cite{ref_OMP} are named as non-NN-based JFSCEs due to the non-NN processing mode and are employed as the baseline for the proposed scheme.

\subsection{Proposed Cascaded ELM-based JFSCE}
Unlike the baseline methods in \cite{ref_ML} and \cite{ref_OMP}, we view the JFSCE as a multi-task problem and develop the cascaded ELM networks to solve this multi-task problem. However, the ELM network is a single-hidden layer network, limiting its learning ability. To tackle this issue, we employ non-NN-based JFSCE as the initial feature extractor for the reason that the initial feature extraction could improve the learning ability of ML-based networks \cite{ELM_fs1,ELM_fs2,OFDM_time_synchron}. The proposed cascaded ELM-based JFSCE is presented in Fig. \ref{cascaded_network}. In this subsection, we first present the initial feature extractor. Then, we elaborate the cascaded ELM networks.

\textit{${\rm{(1)}}$ Initial feature extractor}: As demonstrated in Fig. \ref{cascaded_network}, the initial feature extractors, named non-NN FS and non-NN CE, are employed for the FS and CE, respectively. In this paper, as a demonstration, we employ the non-NN-based JFSCE in \cite{ref_OMP} as the initial feature extractor (Admittedly, other methods can also be applied). For FS, the SM, denoted as ${\bf{u}}_{{\rm{FS}}} \in \mathbb{C} {^{M \times 1}}$, is obtained according to (\ref{sm}), i.e,
\begin{equation}
\label{sm_ini}
{\bf{u}}_{{\rm{FS}}} = {{\bf{S}}^H}{{{\bf{r}}_s^{\left( \tau \right)}}}.
\end{equation}
With the extracted ${\bf{u}}_{{\rm{FS}}}$, we could construct the training data to train the FS-NET, and thus estimate the offset $\widehat \tau$ for FS.

According to the trained FS-NET, we can extract the initial feature for training CE-NET. By denoting the extracted initial feature for CE as ${{\bf{p}}_{{\rm{CE}}}}\ \in \mathbb{C} {^{L \times 1}}$, according to (\ref{h_hat2}), it can be expressed as
\begin{equation}
\label{p_CE}
{{\bf{p}}_{{\rm{CE}}}} = {f_{{\rm{OMP}}}}\left( {{{{\bf{r}}_s^{\left( \tau \right)}}},{{\bf{X}}_\varphi }\left( {\widehat \tau } \right)} \right).
\end{equation}
In (\ref{p_CE}), the ${{\bf{X}}_\varphi }\left( {\widehat \tau } \right) $ is constructed according to $\bf{x}$, i.e., \cite{ref_OMP}
\begin{equation}
\label{eq_13}
{{\bf{X}}_\varphi }\left( {\widehat \tau } \right) = \left[ {\begin{array}{*{20}{c}}
{{x_{M - \widehat \tau }}}&{{x_{M - \widehat \tau - 1}}}& \cdots &{{x_{M - \widehat \tau - L + 1}}}\\
{{x_{M - \widehat \tau + 1}}}&{{x_{M - \widehat \tau }}}& \cdots &{{x_{M - \widehat \tau - L + 2}}}\\
 \vdots & \vdots & \cdots & \vdots \\
{{x_{2M - \widehat \tau - 1}}}&{{x_{2M - \widehat \tau - 2}}}& \cdots &{{x_{2M - \widehat \tau - L}}}
\end{array}} \right].
\end{equation}%
Then, we use the initial feature ${{\bf{p}}_{{\rm{CE}}}}$ to train CE-NET.

With the initial feature extraction, we reap the ${{\bf{u}}_{{\rm{FS}}}}$ and ${\bf{p}}_{{\rm{CE}}}$ for training the cascaded ELM networks. Although
the presented demonstration of the feature extractor above is simply derived from the non-NN-based JFSCE in [5], it is critical for training a single hidden-layer NN, especially for ELM networks.

\textit{${\rm{(2)}}$ Cascaded ELM Networks}: The cascaded ELM network is given in Fig. \ref{cascaded_network}, which is developed to tackle the challenges from the multi-task problem of JFSCE and the influence of HI. The network function, network architecture, and network training and deployment are given as follows.

\textit{${\bf{a)}}$ Network Function Summary}: The cascaded ELM-based JFSCE consists of two subnetworks, named as FS-NET and CE-NET, respectively. The functions of network components are summarized as follows:
\begin{itemize}
  \item FS-NET and CE-NET are cascaded to solve the tasks of FS and CE, respectively.
  \item To alleviate the influence of HI, both FS-NET and CE-NET fuse the non-NN feature extractor and ELM-based networks.
\end{itemize}

\textit{${\bf{b)}}$ Network Architecture}: As shown in Fig. \ref{cascaded_network}, each subnetwork consists of a feature extractor and an ELM network. Each ELM network structures has an input layer, a hidden layer, and an output layer. The network details are given as follows.

\begin{itemize}
  \item The non-NN features extractors, i.e., non-NN FS and non-NN CE, are constructed to extract the initial features for FS-NET and CE-NET, respectively. With the extracted initial features, we fuse the ELM networks to form the FS-NET and CE-NET, respectively. Thereafter, FS-NET and CE-NET are successively cascaded to structure the ELM-based JFSCE framework.
  \item For the ELM network in FS-NET (CE-NET), the neuron numbers of the input layer, hidden layer, and output layer respectively are $M$ ($L$), $\widetilde M = N_mM$ ($\widetilde N = N_lL$), and $M$ ($L$), where ${N_m}$ and $N_l$ are positive integers.
  \item For each ELM network, the sigmoid function, defined as $f \left( x \right) = {1 \mathord{\left/ {\vphantom {1 {\left( {1 + {e^{ - x}}} \right)}}} \right. \kern-\nulldelimiterspace} {\left( {1 + {e^{ - x}}} \right)}}$ \cite{act_function}, is employed as the activation function of the hidden layer, because it is easy to be calculated and commonly used in the shallow neural network \cite{sigmoid_fun}.
 \item The outputs of FS-NET and CE-NET are the offset estimation of FS and CE, respectively.
\end{itemize}

%*******Table1************
\begin{algorithm}[t!]
\SetKwInOut{Input}{Input}
\caption{Cascaded ELM-based JFSCE}
${\bf{Input}}$:
    Received signal: ${{\bf{r}}_s^{\left( \tau  \right)} }$.\\
${\bf{Output}}$: The estimation of FS offset ${\widehat \tau }$, the CE ${{\bf{\widehat h}}}$.\\
\BlankLine
\BlankLine
${\left[ \bf{Offline}\;{\bf{training}\;} \right]}$:\\

Initialize input weight matrix $\bf{W}_{\rm{FS}}$, $\bf{W}_{\rm{CE}}$, and bias vector ${\bf{b}}_{{\rm{FS}}}$ and ${\bf{b}}_{{\rm{CE}}}$.\\
Use (\ref{sm_ini}), (\ref{u_norm})--(\ref{norm_u_FS}) to form the training data-set $\left\{ {\left( {{{\bf{\overline u}}_{{\rm{FS}},i}},{{\bf{t}}_{{\rm{FS}},i}}} \right)} \right\}_{i = 0}^{{N_t}-1}$ by collecting $N_t$ training samples.\\

According to the training data-set $\left\{ {\left( {{{\bf{\overline u}}_{{\rm{FS}},i}},{{\bf{t}}_{{\rm{FS}},i}}} \right)} \right\}_{i = 0}^{{N_t}-1}$, calculate the hidden layer output of ELM in FS-NET (i.e., ${{\bf{o}}_{{\rm{FS,}}i}} $) by using (\ref{o_FS_i}).\\
Collect $N_t$ hidden outputs and training labels to form the hidden layer output matrix ${{{\bf{O}}_{{\rm{FS}}}}}$ and training label matrix ${{{\bf{T}}_{{\rm{FS}}}}}$ according to (\ref{O_FS}) and (\ref{T_FS}), respectively.\\
%\BlankLine
With ${{{\bf{O}}_{{\rm{FS}}}}}$ and $\bf{T}_{\rm{FS}}$, calculate the output weight matrix $\bf{\Omega }_{{\rm{FS}}}$ according to (\ref{Omega_FS}).\\
%\BlankLine
%\BlankLine
Collect $N_c$ training samples to form the training data-set $\left\{ {\left( {{{\bf{\overline p}}_{{\rm{CE}},i}},{{\bf{t}}_{{\rm{CE}},i}}} \right)} \right\}_{i = 0}^{{N_{\rm{c}}}-1}$ by using (\ref{p_CE}), (\ref{p_norm})--(\ref{hat-CE-data-set}).\\
%\BlankLine
Calculate the hidden layer output of ELM in CE-NET (i.e., ${{\bf{o}}_{{\rm{CE}},i}}$) by using ($\ref{o_CE_i}$) according to the training data-set $\left\{ {\left( {{{\bf{\overline p}}_{{\rm{CE}},i}},{{\bf{t}}_{{\rm{CE}},i}}} \right)} \right\}_{i = 0}^{{N_{\rm{c}}}-1}$.\\
%\BlankLine
Use (\ref{O_CE}) and (\ref{T_CE}) to obtain the hidden layer output matrix ${{{\bf{O}}_{{\rm{CE}}}}}$ and the matrix of training label ${{{\bf{T}}_{{\rm{CE}}}}}$ by collecting $N_c$ hidden outputs and training labels.\\
%\BlankLine
Calculate the output weight $\bf{\Omega }_{{\rm{CE}}}$ using (\ref{Omega_CE}) according to ${{{\bf{O}}_{{\rm{CE}}}}}$ and $\bf{T}_{\rm{CE}}$;\\
\BlankLine
\BlankLine
${\left[ \bf{Online}\;{\bf{deployment}\;} \right]}$:\\
%\BlankLine
Extract SM ${{\bf{u}}_{{\rm{FS}}}}$ according to the non-NN FS given in (\ref{sm_ini}) with the received ${{\bf{r}}_s^{\left( \tau  \right)} }$;\\
%\BlankLine
Normalize the SM ${{\bf{u}}_{{\rm{FS}}}}$ as ${{\bf{\overline u}}_{{\rm{FS}}}}$ by using (\ref{u_norm});\\
%\BlankLine
Feed ${\overline{\bf{u}}_{{\rm{FS}}}}$ into the trained FS-NET network to produce the enhanced SM ${\widetilde{\bf{u}}_{{\rm{FS}}}}$ according to (\ref{wave_u_FS}) with the obtained network parameters, i.e., ${{\bf{\Omega }}_{{\rm{FS}}}}$,${\bf{W}_{\rm{FS}}}$, and ${\bf{b}}_{{\rm{FS}}}$;\\
%\BlankLine
Estimate the FS offset to obtain $\widehat \tau$ by using (\ref{tao_hat});\\
%\BlankLine
Extract the initial features of CE (i.e., ${{\bf{ p}}_{{\rm{CE}}}}$) by using the non-NN CE given in (\ref{p_CE}) according to the received ${{\bf{r}}_s^{\left( \tau \right)} }$;\\
%\BlankLine
By using (\ref{p_norm}), normalize the extracted ${{\bf{ p}}_{{\rm{CE}}}}$ to form the normalized initial feature ${{\bf{\overline p}}_{{\rm{CE}}}}$;\\
%\BlankLine
According to the obtained ${{\bf{\overline p}}_{{\rm{CE}}}}$, ${{\bf{\Omega }}_{{\rm{CE}}}}$,${\bf{W}_{\rm{CE}}}$, and ${\bf{b}}_{{\rm{CE}}}$, the enhanced CE ${{\bf{\widehat h}}}$ is estimated by using the trained CE-NET network (which is expressed in (\ref{h_hat})).
\BlankLine
\BlankLine
\end{algorithm}

\textit{${\rm{(3)}}$ Network Training and Deployment}: The network training and deployment are summarized in Algorithm 1, which are elaborated as follows.

\textit{${\bf{a)}}$ Offline Training}: The objects of training the ELM networks of FS-NET and CE-NET are to learn their output weights, respectively. To achieve this multi-task solution, we utilize the subnet-wise training strategy \cite{DeepLearning2NonlinearDis2}. That is, we first train the ELM in FS-NET. Then, the ELM in CE-NET is trained with the trained and fixed network parameters of FS-NET.

\textit{Training ELM Network in FS-NET}: According to (\ref{sm_ini}), $N_t$ SMs  and corresponding training labels are collected to form the data-set
\begin{equation}
\label{data_set}
\left\{ {\left( {{{\bf{u}}_{{\rm{FS}},i}},{{\bf{t}}_{{\rm{FS}},i}}} \right)} \right\}_{i = 0}^{{N_{{t}}-1}},
\end{equation}
where $\left( {{{\bf{u}}_{{\rm{FS}},i}},{{\bf{t}}_{{\rm{FS}},i}}} \right)$ is the $i$-th data sample that consists of $i$-th SM and its training label. In this paper, the SM ${{\bf{u}}_{{\rm{FS}},i}}$ is obtained according to (\ref{sm_ini}). To facilitate the network learning of ELM \cite{ELM_fs2}, we normalize the ${{\bf{u}}_{{\rm{FS}},i}}$ as
\begin{equation}
\label{u_norm}
{{\bf{\overline u}}_{{\rm{FS}},i}} = \frac{{{{\bf{u}}_{{\rm{FS}},i}}}}{{{{\left\| {{{\bf{u}}_{{\rm{FS}},i}}} \right\|}_2}}}.
\end{equation}
In (\ref{data_set}), the training label ${{{\bf{t}}_{{\rm{FS}},i}}}$ is encoded by one-hot coding \cite{ELM_fs1}, which is expressed as
\begin{equation}
\label{t}
{{{\bf{t}}_{{\rm{FS}},i}}} = [\underbrace {0, \cdots ,0}_{{\tau_i}},1,\underbrace {0, \cdots 0}_{M - {\tau_i} - 1}]^T,
\end{equation}
where ${{\tau}_i}$ is the FS offset of the $i$-th data sample. With the normalized SM ${{\bf{\overline u}}_{{\rm{FS}},i}}$, an normalized training data-set is obtained, which is given by
\begin{equation}
\label{norm_u_FS}
\left\{ {\left( {{{\bf{\overline u}}_{{\rm{FS}},i}},{{\bf{t}}_{{\rm{FS}},i}}} \right)} \right\}_{i = 0}^{{N_t}-1}.
\end{equation}
Thus, the training data-set $\left\{ {\left( {{{\bf{\overline u}}_{{\rm{FS}},i}},{{\bf{t}}_{{\rm{FS}},i}}} \right)} \right\}_{i = 1}^{{N_t}} $ is employed to train the ELM in FS-NET. With the training input ${{\bf{\overline u}}_{{\rm{FS}},i}}$, the hidden output of ELM in FS-NET is given by
\begin{equation}
\label{o_FS_i}
{{\bf{o}}_{{\rm{FS,}}i}} = \sigma \left( {{{\bf{W}}_{{\rm{FS}}}}{{\bf{\overline u}}_{{\rm{FS}},i}} + {{\bf{b}}_{{\rm{FS}}}}} \right),
\end{equation}
where ${\bf{W}_{\rm{FS}}} \in \mathbb{C} {^{\widetilde M \times M}}$ and ${\bf{b}}_{{\rm{FS}}} \in \mathbb{C} {^{\widetilde M \times 1 }}$ are the input weight matrix and bias vector, respectively. According to \cite{huang2004extreme} and \cite{huang2006extreme}, the entries of ${\bf{W}_{\rm{FS}}}$ and ${{\bf{b}}_{{\rm{FS}}}}$ are randomly generated. We collect $N_t$ hidden outputs to form the matrix of the hidden layer output as
\begin{equation}
\label{O_FS}
{{\bf{O}}_{{\rm{FS}}}} = \left[ {{{\bf{o}}_{{\rm{FS}}}}_{,0},{{\bf{o}}_{{\rm{FS}}}}_{,1}, \cdots ,{{\bf{o}}_{{\rm{FS}}}}_{,{N_t} - 1}} \right].
\end{equation}
Accordingly, the matrix of training label is constructed as
\begin{equation}
\label{T_FS}
{{\bf{T}}_{{\rm{FS}}}} = [{{\bf{t}}_{{\rm{FS}},0}},{{\bf{t}}_{{\rm{FS}},1}}, \cdots ,{{\bf{t}}_{{\rm{FS}},{N_t} - 1}}].
\end{equation}
With ${{\bf{O}}_{{\rm{FS}}}}$ and ${{\bf{T}}_{{\rm{FS}}}}$, the output weight matrix of ELM in FS-NET, denoted as ${{\bf{\Omega }}_{{\rm{FS}}}} \in \mathbb{C} {^{M \times \widetilde M}}$, is obtained by
\begin{equation}
\label{Omega_FS}
{{\bf{\Omega }}_{{\rm{FS}}}} = {{\bf{T}}_{{\rm{FS}}}}{\bf{O}}_{{\rm{FS}}}^\dag.
\end{equation}
After the ${\bf{\Omega }}_{{\rm{FS}}}$ is trained, the parameters of FS-NET are frozen for training the ELM network in CE-NET.

\textit{Training ELM Network in CE-NET}: According to (\ref{p_CE}), $N_c$ initial features extracted for CE are collected to form the data-set
\begin{equation}
\label{CE-data-set}
\left\{ {\left( {{{\bf{p}}_{{\rm{CE}},i}},{{\bf{t}}_{{\rm{CE}},i}}} \right)} \right\}_{i = 0}^{{N_{\rm{c}}}-1},
\end{equation}
where ${{\bf{p}}_{{\rm{CE}},i}}$ and ${{\bf{t}}_{{\rm{CE}},i}}$ are the $i$-th extracted initial feature and its corresponding training label, respectively. For the benefit of the ELM training, ${{\bf{p}}_{{\rm{CE}},i}}$ is normalized as
\begin{equation}
\label{p_norm}
{{\bf{\overline p}}_{{\rm{CE}},i}} = \frac{{{{\bf{p}}_{{\rm{CE}},i}}}}{{{{\left\| {{{\bf{p}}_{{\rm{CE}},i}}} \right\|}_2}}}.
\end{equation}
In (\ref{CE-data-set}), the training label ${{\bf{t}}_{{\rm{CE}},i}} \in \mathbb{C} {^{L \times 1}}$ is generated according to \cite{rician_fading} and (\ref{eq_3}). With ${{\bf{\overline p}}_{{\rm{CE}},i}}$, the normalized training data-set is given by
\begin{equation}
\label{hat-CE-data-set}
\left\{ {\left( {{{\bf{\overline p}}_{{\rm{CE}},i}},{{\bf{t}}_{{\rm{CE}},i}}} \right)} \right\}_{i = 0}^{{N_{\rm{c}}}-1}.
\end{equation}
By using ${{\bf{\overline p}}_{{\rm{CE}},i}}$ as the training input, the hidden output of the ELM in CE-NET is obtained, which is expressed as
\begin{equation}
\label{o_CE_i}
{{\bf{o}}_{{\rm{CE}},i}} = \sigma \left( {{{\bf{W}}_{{\rm{CE}}}}{{\bf{\overline p}}_{{\rm{CE}},i}} + {{\bf{b}}_{{\rm{CE}}}}} \right),
\end{equation}
where ${\bf{W}_{\rm{CE}}} \in \mathbb{C} {^{\widetilde N \times L}}$ and bias ${\bf{b}}_{{\rm{CE}}} \in \mathbb{C} {^{\widetilde N \times 1 }}$ are weight matrix and bias vector, respectively. The entries of ${\bf{W}_{\rm{CE}}}$ and ${{\bf{b}}_{{\rm{CE}}}}$ are randomly generated for parameter initialization \cite{huang2004extreme} and \cite{huang2006extreme}. Then, $N_c$ hidden outputs are collected to form the hidden output matrix ${{\bf{O}}_{{\rm{CE}}}} \in \mathbb{C} {^{\widetilde N \times {N_c} }} $ as
 \begin{equation}
\label{O_CE}
{{\bf{O}}_{{\rm{CE}}}} = \left[ {{{\bf{o}}_{{\rm{CE}},0}},{{\bf{o}}_{{\rm{CE}},1}}, \cdots ,{{\bf{o}}_{{\rm{CE}},{N_c} - 1}}} \right].
\end{equation}
 Likewise, the $N_c$ collected ${{\bf{t}}_{{\rm{CE}},i}}$ are represented as the matrix
  \begin{equation}
\label{T_CE}
{{\bf{T}}_{{\rm{CE}}}} = [{{\bf{t}}_{{\rm{CE}},0}},{{\bf{t}}_{{\rm{CE}},1}}, \cdots ,{{\bf{t}}_{{\rm{CE}},{N_c} - 1}}].
\end{equation}
Then, the output weight matrix of ELM in CE-NET (denoted by ${{\bf{\Omega }}_{{\rm{CE}}}} \in \mathbb{C} {^{L \times \widetilde N}}$) is obtained, which is given by
\begin{equation}
\label{Omega_CE}
{{\bf{\Omega }}_{{\rm{CE}}}} = {{\bf{T}}_{{\rm{CE}}}}{\bf{O}}_{{\rm{CE}}}^\dag.
\end{equation}

With the trained ${\bf{\Omega }}_{{\rm{FS}}}$ in (\ref{Omega_FS}) and ${\bf{\Omega }}_{{\rm{CE}}}$ in (\ref{Omega_CE}), network parameters of FS-NET and CE-NET are obtained, and thus the online deployment can be performed.

\textit{${\bf{b)}}$ Online Deployment}: The main task of cascaded ELM-based JFSCE is to obtain an enhanced estimation of FS offset $\widehat \tau$ and CE ${{\bf{\widehat h}}}$. The online deployment is implemented as follows. Firstly, the feature extractor with non-NN FS in (\ref{sm_ini}) is employed to extract the initial feature ${{\bf{u}}_{{\rm{FS}}}}$ and its normalized version ${{\bf{\overline u}}_{{\rm{FS}}}}$ is obtained by using (\ref{u_norm}). Based on ${{\bf{\overline u}}_{{\rm{FS}}}}$, the FS-NET produces an enhanced SM ${\widetilde{\bf{u}}_{{\rm{FS}}}} \in \mathbb{C} {^{M \times 1}}$ as
\begin{equation}
\label{wave_u_FS}
{\widetilde{\bf{u}}_{{\rm{FS}}}} = {{\bf{\Omega }}_{{\rm{FS}}}}\left( {{{\bf{W}}_{{\rm{FS}}}}{{\bf{\overline u}}_{{\rm{FS}}}} + {{\bf{b}}_{{\rm{FS}}}}} \right).
\end{equation}
By denoting ${\widetilde{\bf{u}}_{{\rm{FS}}}} = {\left[ {{\widetilde u_{{\rm{FS}},0}},{\widetilde u_{{\rm{FS}},1}}, \cdots ,{\widetilde u_{{\rm{FS}},M - 1}}} \right]^T}$, the estimation of FS offset (i.e., $\widehat \tau$) is given by
\begin{equation}
\label{tao_hat}
\widehat \tau =\mathop{\arg \max} \limits_{0 \le i \le M - 1} {\left| {{\widetilde u_{{\rm{FS}},i}}} \right|^2}.
\end{equation}
With the estimated $\widehat \tau$, we can determine the start point of a frame, and thus extract the received training sequence from the received signal to perform CE. According to (\ref{p_CE}), the non-NN CE is utilized to extract the initial feature ${{\bf{p}}_{{\rm{CE}}}}$. Then, the normalized ${{\bf{\overline p}}_{{\rm{CE}}}}$ is obtained from (\ref{p_norm}). Based on ${{\bf{\overline p}}_{{\rm{CE}}}}$, the CE-NET is performed to enhance CE, which is expressed as
\begin{equation}
\label{h_hat}
{\bf{\widehat h}} = {{\bf{\Omega }}_{{\rm{CE}}}}\left( {{{\bf{W}}_{{\rm{CE}}}}{{\bf{\overline p}}_{{\rm{CE}}}} + {{\bf{b}}_{{\rm{CE}}}}} \right).
\end{equation}

According to (\ref{wave_u_FS})-(\ref{h_hat}), with the normalized versions of initial features, i.e., ${{\bf{\overline u}}_{{\rm{FS}}}}$ and ${{\bf{\overline p}}_{{\rm{CE}}}}$, the cascaded ELM networks are performed, promoting the FS-NET and CE-NET to jointly enhance the FS and CE. Especially, due to the cascaded ELM networks, the nonlinear influence of HI is effectively alleviated in both phases of FS and CE. The exploited strategy fuses non-NN and NN modes to deal with the nonlinear influence, forming a paradigm from a new perspective of fusion learning for JFSCE.

%=======================第四章 实验仿真=============================================
\section{NUMERICAL SIMULATION}
In this section, numerical simulations are performed to evaluate the effectiveness and robustness of the proposed cascaded ELM-based JFSCE method. During simulations, the proposed method is compared with the classical JFSCE in \cite{ref_ML} and the recent orthogonal matching pursuit (OMP)-based algorithm in \cite{ref_OMP} in Rician fading channel scenarios with HI. The basic parameter setting involved is listed below. The training sequence is Zadoff-Chu sequence \cite{c20}, ${N_s} = 32$, ${M = 160}$, ${\widetilde M = 10M = 1600}$, ${N_t} = 1 \times {10^5}$, $L = 8$, and $K = 8$. The data sequence $\bf{d}$ in the transmitted frame is modulated by quadrature-phase shift-keying (QPSK). The signal-to-noise ratio (SNR) in decibel (dB) is defined as ${\rm{SNR}} = 10{\log _{10}}\left( {{P \mathord{\left/ {\vphantom {P {{\sigma ^2}}}} \right.\kern-\nulldelimiterspace} {{\sigma ^2}}}} \right)$ \cite{DeepLearning2NonlinearDis2}. And the exponentially decayed power coefficient ${\eta}$ of the Rician fading channel is set as $0.2$. The performance of FS and CE in JFSCE are evaluated by the error probability and NMSE, respectively. In the simulations, the error probability of FS is defined as
\begin{equation}
\label{e_error}
{e_{{\rm{error}}}} = \frac{{{N_{{\rm{error}}}}}}{{{N_r}}},
\end{equation}
where ${{N_{{\rm{error}}}}}$ denotes the times of error FS in $N_r$ experiments. The NMSE of CE is defined as
\begin{equation}
\label{epsilon_NMSE}
{\overline \varepsilon _{{\rm{NMSE}}}} = \mathbb{E} \left \{ {\frac{{\left\| {{{{\bf{\widehat h}}}} - {\bf{h}}} \right\|_2^2}}{{\left\| {\bf{h}} \right\|_2^2}}} \right \},
\end{equation}
where $\bf{\widehat h}$ denotes the estimation of $\bf{h}$.
For HI, the effects of nonlinear distortion caused by HPA are considered. According to the HPA given in \cite{EVM_formula}, the nonlinear amplitude $A(x)$ and phase $\Phi \left( x \right)$ are defined as
\begin{equation}
\label{eq_QQQQQQ}
\left\{ {\begin{array}{*{20}{c}}
{A\left( x \right) = \frac{{{\alpha _a}x}}{{\left( {1 + {\beta _a}{x^2}} \right)}}}\\
{\Phi \left( x \right) = \frac{{{\alpha _\varphi }{x^2}}}{{\left( {1 + {\beta _\varphi }{x^2}} \right)}}}
\end{array}} \right.,
\end{equation}
where ${\alpha _a} = 2.16$, ${\beta _a} = 1.15$, ${\alpha _\varphi } = 4.00$, and ${\beta _\varphi } = 9.10$ are set to reflect the impact of HI on the signal \cite{EVM_formula}. We employ the error vector magnitude (EVM) to evaluate the distortion intensity, which is defined as \cite{TWT_params}
\begin{equation}
 \label{eq_32}
{\delta _{{\rm{EVM}}}}\left( {\rm{\% }} \right)= \sqrt {\frac{{{{\left\| {{{{\bf{\widetilde x}}}_n} - {{\bf{x}}_{\mathrm{ref}}}} \right\|}_2}}}{{{{\left\| {{{\bf{x}}_{\mathrm{ref}}}} \right\|}_2}}}},
\end{equation}
where ${{{\bf{\widetilde x}}}_n}$ represents the distorted signal through HPA. To maximize the power efficiency, the HPA is considered to work in its saturation region \cite{EVM_saturation}. In (\ref{eq_32}), ${{{\bf{x}}_{{\rm{ref}}}}}$ is the reference signal vector and denotes the received symbol vector linearly amplified by HPA (without distortion). In this paper, the EVM is set as ${\delta _{{\rm{EVM}}}} = 35{\rm{\% }}$ for the basic parameter setting.

%*******Table2 DNN,ELM_Learn和Prop复杂度比较************
\textcolor[rgb]{0.00,0.07,1.00}{\begin{table*}[htbp]
  \vspace{-3mm}
\renewcommand{\arraystretch}{1.25}
\caption{Testing Computational Complexity among Different JFSCE Methods}
\label{table_II}
\centering
\scriptsize
\setlength{\tabcolsep}{2mm}{
\begin{tabu}{@{}c|c|c@{}}
\tabucline[1pt]{-}
    Method          & CM   & Examples \\
 \tabucline[1pt]{-}%\Xhline{0.8pt}
    DNN-based            & $\left( {\frac{1}{2}{N_{\mathrm{DNN},1}} + 1} \right){M^2} + {L^2}M + \frac{1}{2}{N_{\mathrm{DNN},2}}{L^2} + \sum\limits_{\iota = 1}^L {2 \iota M + 2{{\iota}^2}M + {{\iota}^3}}$                                                             &  $627216$              \\ \hline
    ELM\_Learn     & $2{N_{\mathrm{ELM},1}}{M^2} +\left({N_{\mathrm{ELM},2}}ML + {N_{\mathrm{ELM},2}}L^2 \right)$                  &  $630528$        \\ \hline
    {Prop}     & $\left( {2{N_m} + 1} \right){M^2} + \left( {{L^2}M + \sum\limits_{\iota = 1}^L {{2\iota M} + 2{{\iota}^2}M + {{\iota}^3}} } \right) + 2{N_l}{L^2}$
                                                                                           &  $627216$             \\
    \tabucline[1pt]{-}
\end{tabu}}
  \vspace{-3mm}
\end{table*}}

For expression convenience, the ``Prop'', ``Ref\_\hspace{0.01em}\cite{ref_ML}'', ``Ref\_\hspace{0.01em}\cite{ref_OMP}'', ``ELM\_learn'', and ``DNN-based'' are used to denote the proposed cascaded ELM-based JFSCE, the JFSCE in \cite{ref_ML}, the OMP-based JFSCE in \cite{ref_OMP}, the ELM-based JFSCE method without preprocessing and the DNN-based method in \cite{ref_DNN}, respectively. For fair comparison, a cascaded DNN-based structure is also employed for the ``DNN-based''. The first subnetwork of ``DNN-based'' consists of a non-NN feature extractor and a single-hidden-layer NN, where the non-NN feature extractor is the same as that of FS-Net. For the single-hidden-layer NN in the first subnetwork of ``DNN-based'', the neurons of the input layer, hidden layer, and output layer are set as $M$, ${{N_{\mathrm{DNN},1}}}M$, and $M$, respectively. Similarly, the second subnetwork of ``DNN-based'' consists of the non-NN feature extractor and the single-hidden-layer NN as well. Its non-NN feature extractor is the same as that of CE-Net, and the single-hidden-layer NN owns the neurons of the input layer, hidden layer, and output layer $L$, ${{N_{\mathrm{DNN},2}}}L$, and $L$, respectively.

\subsection{Computational Complexity}
The comparison of computational complexity of NN-based methods is summarized in TABLE I, in which complex multiplication (CM) is employed to measure the computational complexity \cite{multiply_complex}. For ``DNN-based'', the CMs are $\left( {\frac{1}{2}{N_{\mathrm{DNN},1}} + 1} \right){M^2} + {L^2}M + \frac{1}{2}{N_{\mathrm{DNN},2}}{L^2} + \sum\limits_{\iota = 1}^L {\left ({2\iota M} + 2{{\iota}^2}M + {{\iota}^3}\right )}$, which mainly reflects two parts given as follows. 1) For the non-NN feature extractors in the first and the second subnetworks of ``DNN-based'', the CMs are $\frac{1}{2}{N_{\mathrm{DNN},1}}{M^2}$ and ${{L^2}M + \sum\limits_{\iota = 1}^L \left({{2\iota M} + 2{{\iota}^2}M + {{\iota}^3}}\right )} $ (according to \cite{ref_OMP}), respectively. 2) For the single-hidden-layer NN, ${M^2}$ and $\frac{1}{2}{N_{\mathrm{DNN},2}}{L^2}$ CMs (according to \cite{FC_complex}) are respectively computed in the first and the second subnetworks of ``DNN-based''. Similarly, the CMs of non-NN FS and non-NN CE are ${2{N_m}}M^2$ and ${{L^2}M + \sum\limits_{\iota = 1}^L \left ( {{2\iota M} + 2{{\iota}^2}M + {{\iota}^3}}\right) }$, respectively. For the ELM networks in FS-NET and CE-NET, ${M^2}$ and $2{N_l}{L^2}$ CMs are respectively computed. Then, the total CMs of ``Prop'' are $\left( {2{N_m} + 1} \right){M^2} + {L^2}M + 2{N_l}{L^2} + \sum\limits_{\iota = 1}^L \left ({{2\iota M} + 2{{\iota}^2}M + {{\iota}^3}}\right)$. For ``ELM\_learn'', the neurons of hidden layers of two ELM networks are respectively set as ${N_{\mathrm{ELM},1}}M$ and ${N_{\mathrm{ELM},2}}L$. Here, due to the lack of preprocessing in ``ELM\_learn'', ${N_{\mathrm{ELM},1}}$ and ${N_{\mathrm{ELM},2}}$ are usually larger than ${N_{m}}$ and ${N_{l}}$, respectively. The total CMs of ``ELM\_learn'' are $2{N_{\mathrm{ELM},1}}{M^2} +\left({N_{\mathrm{ELM},2}}ML + {N_{\mathrm{ELM},2}}L^2 \right)$, in which the ELM networks of its first and second subnetwork have $2{N_{\mathrm{ELM},1}}{M^2}$ and ${N_{\mathrm{ELM},2}}{M^2} + {N_{\mathrm{ELM},2}}ML$ CMs, respectively. An example is also given in TABLE I, where ${M = 160}$, $N_m = 10$, $L = 8$, ${N_{\mathrm{DNN},1}}{\rm{ = }}{N_{\mathrm{DNN},2}}{\rm{ = }}40$ (with the same computational complexity as that of ``Prop''), and ${N_{\mathrm{ELM},1}}{\rm{ = }}{N_{\mathrm{ELM},2}}{\rm{ = }}12$ (with a higher computational complexity than that of ``Prop'') are considered. From TABLE I, the proposed JFSCE method owns the same computational complexity as that of ``DNN-based''. Compared with ``ELM\_Learn'', a lower computational complexity is achieved by the proposed JFSCE method.

With the same or lower computational complexity relative to ``DNN-based'' and ``ELM\_Learn'', we validate that the FS and CE performance can be improved in Section IV-B.

%the effectiveness and robustness of the cascaded ELM-based JFSCE \textcolor[rgb]{1.00,0.00,0.00}{are elaborated in Section IV-B and IV-C, respectively.

%
\begin{figure}[h]
    \includegraphics[width=0.9\columnwidth]{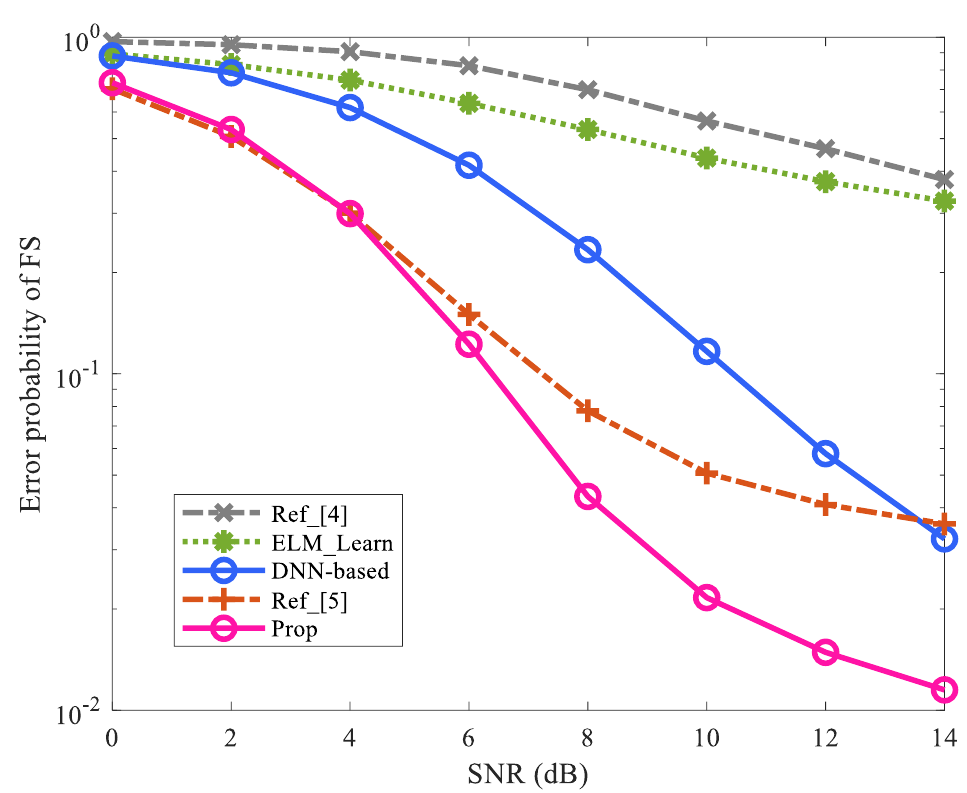}
    \caption{Error probability of FS, where ${\delta _{{\rm{EVM}}}} = 35{\rm{\% }}$, $K=8$, $L=8$, $N_s=32$, and $M=160$.}
    \label{performanceFS}
    \centering
    \includegraphics[width=0.9\columnwidth]{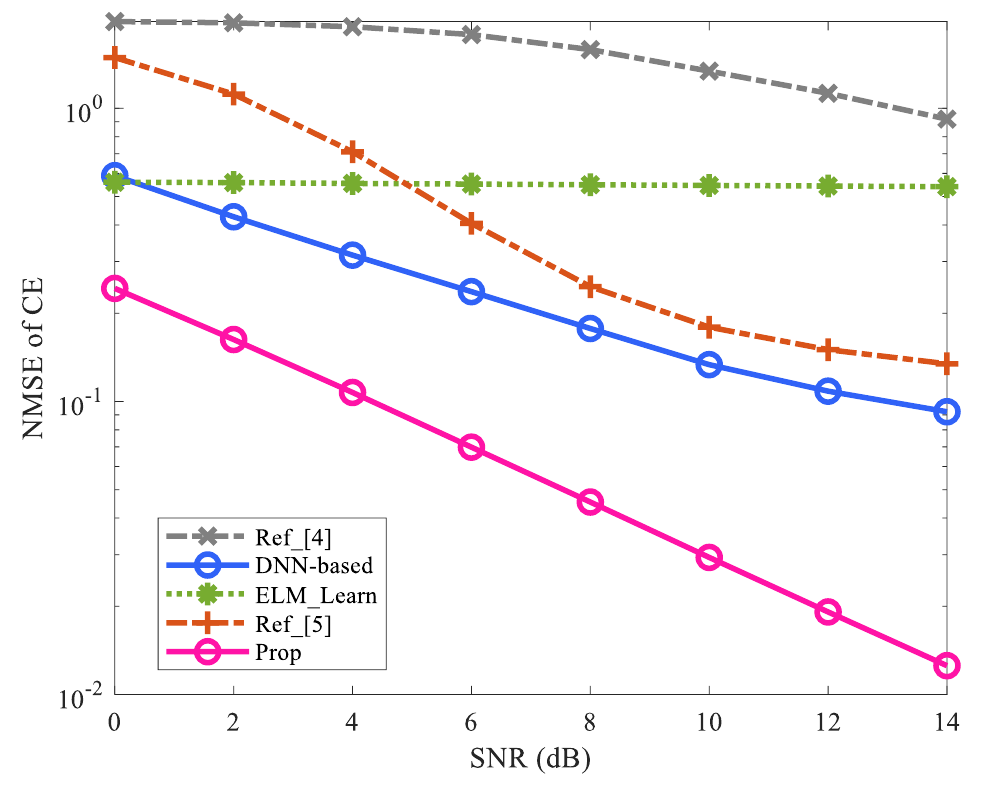}
    \caption{NMSE of CE, where ${\delta _{{\rm{EVM}}}} = 35{\rm{\% }}$, $K=8$, $L=8$, $N_s=32$, and $M=160$.}
    \label{performanceCE}
\end{figure}

\text{\\ \\ \\}

\subsection{Effectiveness of the cascaded ELM-based JFSCE}
%---------A有效性-------
To illustrate the effectiveness of the cascaded ELM-based JFSCE, we compare ``Prop'' with other four methods (i.e., ``Ref\_\hspace{0.01em}\cite{ref_ML}'', ``Ref\_\hspace{0.01em}\cite{ref_OMP}'', ``ELM\_learn'', and ``DNN-based'') according to the error probability of FS (i.e., ${e_{{\rm{error}}}}$) and the NMSE (i.e., ${\overline \varepsilon _{{\rm{NMSE}}}}$) of CE in Fig. \ref{performanceFS} and Fig. \ref{performanceCE}, respectively. From Fig. \ref{performanceFS}, the error probability of ``Prop'' is significantly lower than those of ``Ref\_\hspace{0.01em}\cite{ref_ML}'', ``Ref\_\hspace{0.01em}\cite{ref_OMP}'', and ``ELM\_learn'' when ${\rm{SNR}}$ $\geq6$dB. For example, when ${\rm{SNR}}$ = 10dB, the value of ${e_{{\rm{error}}}}$ for ``Prop'' is about $2.22 \times {10^{{\rm{ - }}2}}$, while the values of ${e_{{\rm{error}}}}$ are about $5.65 \times {10^{{\rm{ - }}1}}$, $5.36 \times {10^{{\rm{ - }}2}}$, and $4.57 \times {10^{{\rm{ - }}1}}$ for ``Ref\_\hspace{0.01em}\cite{ref_ML}'', ``Ref\_\hspace{0.01em}\cite{ref_OMP}'', and ``ELM\_learn'', respectively. The FS-NET significantly improves the error probability of FS. The reason is that the ELM in FS-NET effectively alleviates the influence of HI, while the other methods of JFSCE encounter severe performance degradation due to the lack of consideration for HI. Meanwhile, the values of ${e_{{\rm{error}}}}$ for ``Prop'' are significantly smaller than that of ``ELM\_learn'' in the whole given SNR region. Especially in the relatively high SNR region, e.g., ${\rm{SNR}}$ $\geq4$dB, the curve gap between the ``Prop'' and ``ELM\_learn'' becomes larger as SNR increases. This implies that the non-NN FS plays an effective role in ELM learning due to its structure of the single hidden layer (limiting its learning ability). In addition, the error probability of the ``Prop'' is lower than that of the ``DNN-based''. This proves that, compared with the ``DNN-based'', the ``Prop'' significantly improves the FS performance with approximate computational complexity. Thus, to alleviate the influence of HI, it is an effective FS mode by fusing the non-NN FS and ELM.

Fig. \ref{performanceCE} validates the effectiveness of the cascaded ELM-based JFSCE in NMSE performance. From Fig. \ref{performanceCE}, the NMSE of ``Prop'' is significantly smaller than those of the other four schemes. For example, when ${\rm{SNR}}$ = 6dB, the NMSE of ``Prop'' is about $7.03 \times {10^{{\rm{ - }}2}}$, while the NMSEs of ``Ref\_\hspace{0.01em}\cite{ref_ML}'', ``Ref\_\hspace{0.01em}\cite{ref_OMP}'', ``ELM\_learn'', and ``DNN-based'' are all higher than $1.00 \times {10^{{\rm{ - }}1}}$. The NMSE performance of the proposed JFSCE significantly outperforms those of ``Ref\_\hspace{0.01em}\cite{ref_ML}'', ``Ref\_\hspace{0.01em}\cite{ref_OMP}'', ``ELM\_learn'', and ``DNN-based''. One reason is that the ``Prop'' benefits from the architecture of cascaded ELM networks. The correctness improvement of FS (by using FS-NET) naturally promotes the following CE-NET to improve its CE accuracy. Besides, the CE-NET fuses the non-NN and NN-based CE, and thus possesses a good ability to solve the nonlinear problems caused by HI.

On the whole, from Fig. \ref{performanceFS} and Fig. \ref{performanceCE}, both the FS's error probability and CE's NMSE are improved compared with ``Ref\_\hspace{0.01em}\cite{ref_ML}'', ``Ref\_\hspace{0.01em}\cite{ref_OMP}'', ``ELM\_learn'', and ``DNN-based''. Thus, the proposed cascaded ELM-based JFSCE possesses its effectiveness in the scenarios of the Rician fading channel with HI.

\begin{figure}[h]
\centering    %居中
\subfigure[Error probability of FS] %第一张子图
{
%	\begin{minipage}{0.48\textwidth}
	\centering        %子图居中
	\includegraphics[scale=0.8]{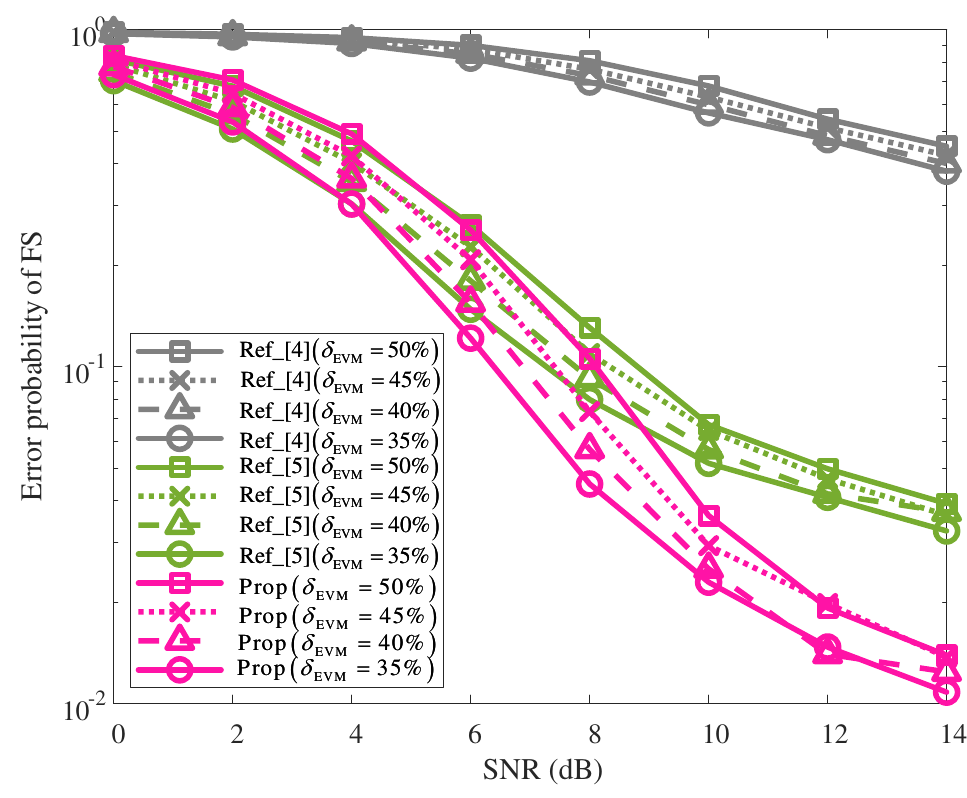}   %以pic.jpg的0.5 倍大小输出
%	\end{minipage}
	%\label{JFSCE_Vs_KKFS}
}	
\subfigure[NMSE of CE] %第二张子图
{
%	\begin{minipage}{0.48\textwidth}
	\centering      %子图居中
	\includegraphics[scale=0.8]{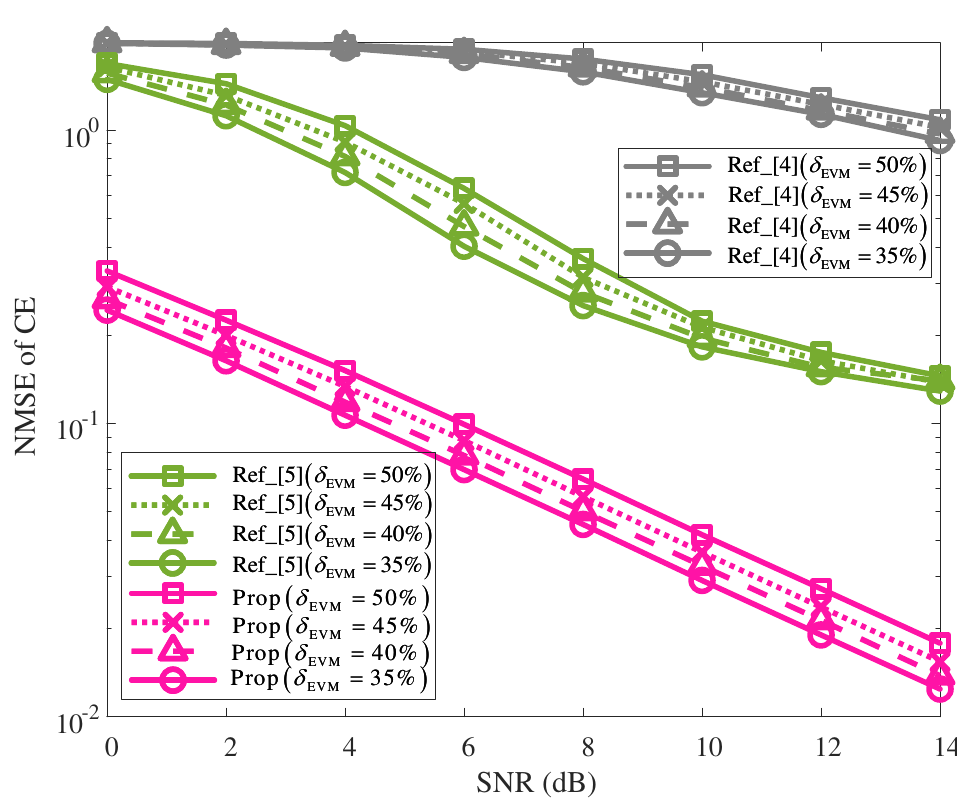}   %以pic.jpg的0.5 倍大小输出
	%\end{minipage}
	%\label{JFSCE_Vs_KKCE}  %图片引用标记
}
\caption{Error probability of FS and NMSE of CE against the impact of ${\delta _{{\rm{EVM}}}}$, where ${\delta _{{\rm{EVM}}}}{\rm{ = }}35{\rm{\% }}$, ${\delta _{{\rm{EVM}}}}{\rm{ = }}40{\rm{\% }}$, ${\delta _{{\rm{EVM}}}}{\rm{ = }}45{\rm{\% }}$, and ${\delta _{{\rm{EVM}}}}{\rm{ = }}50{\rm{\% }}$.} %  % 大图名称
\label{JFSCE_Vs_EVMFSandCE}
\end{figure}

%--------B 鲁棒性---------
\subsection{Robustness Analysis}
\begin{figure}[h]
\centering    %居中
\subfigure[Error probability of FS] %第一张子图
{
	\begin{minipage}{0.48\textwidth}
	\centering          %子图居中
	\includegraphics[scale=0.8]{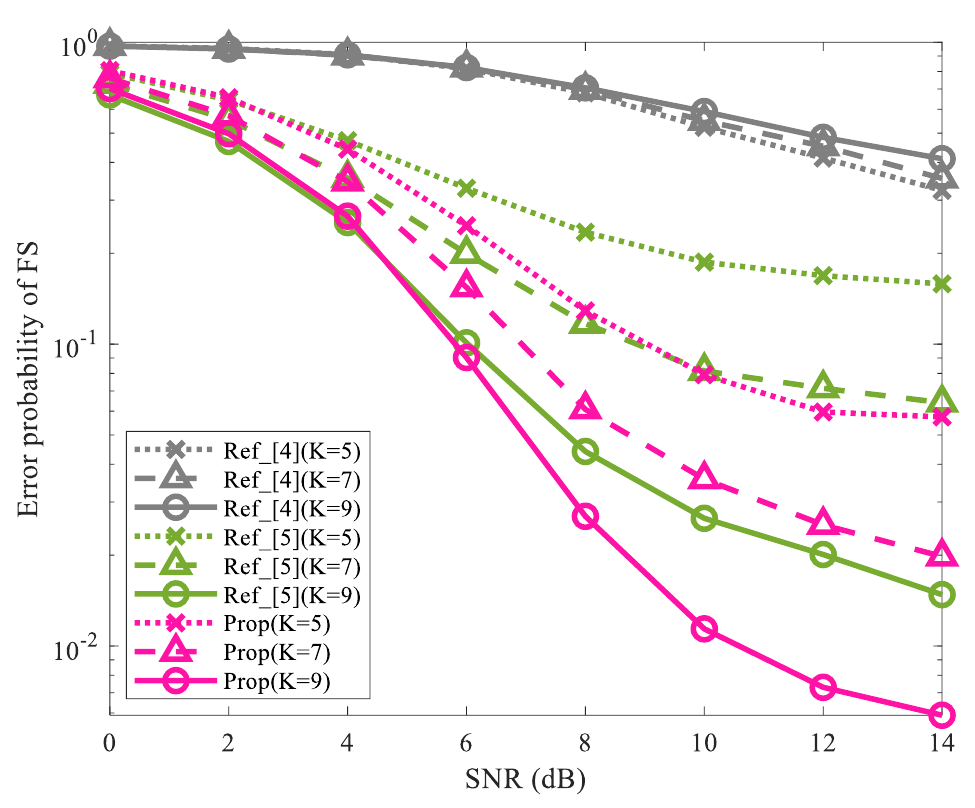}   %以pic.jpg的0.5倍大小输出
	\end{minipage}
	%\label{JFSCE_Vs_KKFS}
}	
\subfigure[NMSE of CE] %第二张子图
{
	\begin{minipage}{0.48\textwidth}
	\centering      %子图居中
	\includegraphics[scale=0.8]{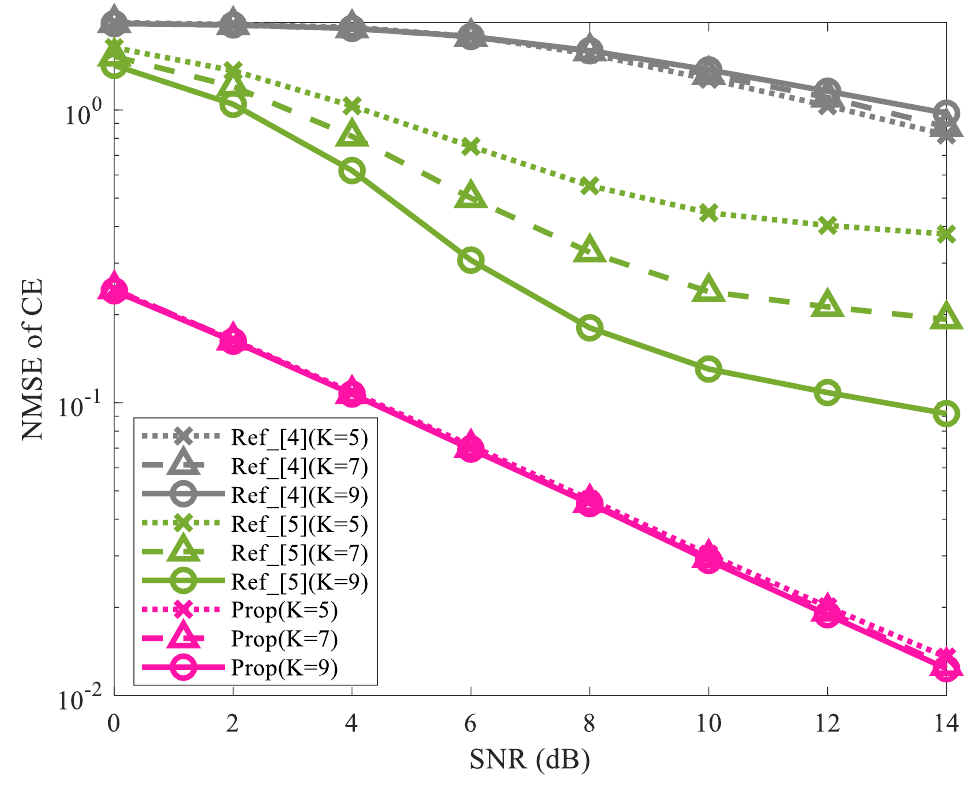}   %以pic.jpg的0.5倍大小输出
	\end{minipage}
	%\label{JFSCE_Vs_KKCE}  %图片引用标记
}
\caption{Error probability of FS and NMSE of CE against the impact of $K$, where $K=5$, $K=7$, and $K=9$.} %  %大图名称
\label{JFSCE_Vs_KKFSandCE}
\end{figure}

To evaluate the robustness of the cascaded ELM-based JFSCE, simulations are given to demonstrate the impacts against the EVM, the Rician $K$ factor, the number of multi-path $L$, the length of training sequence $N_s$, and the length of observation window $M$ from Fig. \ref{JFSCE_Vs_EVMFSandCE} to Fig. \ref{JFSCE_Vs_NNFSandCE}. For ease of analysis, except for the change of impact parameters (i.e., ${\delta _{{\rm{EVM}}}}$, $K$, $L$, $N_s$, and $M$), other basic parameters remain the same as those in Fig. \ref{performanceFS} and Fig. \ref{performanceCE} during the simulations.

\subsubsection{Robustness against ${\delta _{{\rm{EVM}}}}$}
To demonstrate the impact of EVM, Fig.\ref{JFSCE_Vs_EVMFSandCE} shows the FS's error probability and CE's NMSE, where ${\delta _{{\rm{EVM}}}}{\rm{ = }}35{\rm{\% }}$, ${\delta _{{\rm{EVM}}}}{\rm{ = }}40{\rm{\% }}$, ${\delta _{{\rm{EVM}}}}{\rm{ = }}45{\rm{\% }}$, and ${\delta _{{\rm{EVM}}}}{\rm{ = }}50{\rm{\% }}$ are considered. From Fig. \ref{JFSCE_Vs_EVMFSandCE} (a), it is observed that, with the value of ${\delta _{{\rm{EVM}}}}$ increases, the error probability of FS increases due to the increased distortion intensity. Although the error probability of FS for each given method in Fig. \ref{JFSCE_Vs_EVMFSandCE} (a) increases, the proposed method almost achieves the smallest error probability in all given SNR regions. For given EVM and SNR, e.g, ${\delta _{{\rm{EVM}}}}{\rm{ = }}40{\rm{\% }}$ and ${\rm{SNR}}$ = 8dB, the ``Prop'' attains an error probability of $5.68 \times {10^{{\rm{ - }}2}}$, while the error probabilities of ``Ref\_\hspace{0.01em}\cite{ref_ML}'' and ``Ref\_\hspace{0.01em}\cite{ref_OMP}'' are about $7.30 \times {10^{{\rm{ - }}1}}$ and $9.22 \times {10^{{\rm{ - }}2}}$, respectively. This indicates that, even against the varying of EVM, the ELM in FS-NET still effectively alleviates the nonlinear influences caused by HI, and thus obtains the smallest error probability. From Fig. \ref{JFSCE_Vs_EVMFSandCE} (b), with the increase of EVM (from 35\% to 50\% with the interval of 5\%), the NMSE of each given method (i.e., the CE methods of ``Prop'', ``Ref\_\hspace{0.01em}\cite{ref_ML}'', and ``Ref\_\hspace{0.01em}\cite{ref_OMP}'') increases. For example, when ${\rm{SNR}}$ = 8dB and EVMs are from 35\% to 50\%, the values of NMSE of the ``Ref\_\hspace{0.01em}\cite{ref_OMP}'' are $2.53 \times {10^{{\rm{ - }}1}}$, $2.80 \times {10^{{\rm{ - }}1}}$, $3.15 \times {10^{{\rm{ - }}1}}$, and $3.65 \times {10^{{\rm{ - }}1}}$, respectively. Even so, the NMSE of the ``Prop'' is significantly smaller than those of ``Ref\_\hspace{0.01em}\cite{ref_ML}'' and ``Ref\_\hspace{0.01em}\cite{ref_OMP}''. For a given EVM and SNR, e.g., ${\delta _{{\rm{EVM}}}}{\rm{ = }}45{\rm{\% }}$ and ${\rm{SNR}}$ = 10dB, the NMSE of ``Ref\_\hspace{0.01em}\cite{ref_ML}'' and ``Ref\_\hspace{0.01em}\cite{ref_OMP}'' are larger than $1.00 \times {10^{{\rm{ - }}1}}$, while the NMSE of the ``Prop'' is lower than $5.00 \times {10^{{\rm{ - }}2}}$. This reflects that the CE is significantly improved by ``Prop'', which benefits from the architecture of cascaded FS-ELM and CE-NET, the ability to solve the nonlinear influences by using ELM in CE-NET, and the strategy of fusing non-NN and NN-based CE. Against the impact of EVM, from Fig. \ref{JFSCE_Vs_EVMFSandCE} (a) and Fig. \ref{JFSCE_Vs_EVMFSandCE} (b), the proposed cascaded ELM-based JFSCE presents smaller values of FS's error probability and CE's NMSE than those of ``Ref\_\hspace{0.01em}\cite{ref_ML}'' and ``Ref\_\hspace{0.01em}\cite{ref_OMP}'', and thus obtains excellent robustness.

\subsubsection{Robustness against $K$}
To illustrate the impact of varying Rician factor $K$, the error probability of FS and the NMSE of CE are plotted in Fig. \ref{JFSCE_Vs_KKFSandCE}(a) and Fig. \ref{JFSCE_Vs_KKFSandCE}(b), respectively. In Fig. \ref{JFSCE_Vs_KKFSandCE}, $K = 5, K = 7$, and $K = 9$ are respectively considered. From Fig. \ref{JFSCE_Vs_KKFSandCE}(a), the FS's error probability decreases with the increasing value of $K$ due to the increased possibility of the LOS component. For example, for ${\rm{SNR}}$ = 8dB, when $K = 5, K = 7$, and $K = 9$, the error probabilities of ``Ref\_\hspace{0.01em}\cite{ref_OMP}'' are about $2.35 \times {10^{{\rm{ - }}1}}$, $1.17 \times {10^{{\rm{ - }}1}}$, and $4.41 \times {10^{{\rm{ - }}2}}$, respectively. For each given $K$, the FS's error probability of the ``Prop'' is minimum in the given SNR region (from 0dB to 14dB). For example, when $K = 7$ and ${\rm{SNR}}$ = 10dB, the error probability of ``Prop'' attains ${e_{{\rm{error}}}}$ = $3.57 \times {10^{{\rm{ - }}2}}$, while those of ``Ref\_\hspace{0.01em}\cite{ref_ML}'' and ``Ref\_\hspace{0.01em}\cite{ref_OMP}'' are about $5.49 \times {10^{{\rm{ - }}1}}$ and $8.15 \times {10^{{\rm{ - }}2}}$, respectively. This reflects that, no matter whether the Rician factor $K$ increases or decreases, the ``Prop'' significantly decreases the FS's error probability when compared with ``Ref\_\hspace{0.01em}\cite{ref_ML}'' and ``Ref\_\hspace{0.01em}\cite{ref_OMP}''. From Fig. \ref{JFSCE_Vs_KKFSandCE}(b), for different values of $K$, the ``Prop'' achieves the minimal NMSE. For $K = 5$ and ${\rm{SNR}}$ = 8dB, ``Prop'' achieves ${\overline \varepsilon _{{\rm{NMSE}}}}$ = $4.67 \times {10^{{\rm{ - }}2}}$, while the NMSEs of ``Ref\_\hspace{0.01em}\cite{ref_ML}'' and ``Ref\_\hspace{0.01em}\cite{ref_OMP}'' are about $1.55$ and $5.51 \times {10^{{\rm{ - }}1}}$, respectively. This embodies that, for the varying Rician factor $K$, the ``Prop'' can still achieve the minimum of NMSE compared with ``Ref\_\hspace{0.01em}\cite{ref_ML}'' and ``Ref\_\hspace{0.01em}\cite{ref_OMP}''. On the whole, from Fig. \ref{JFSCE_Vs_KKFSandCE}(a) and Fig. \ref{JFSCE_Vs_KKFSandCE}(b), the smallest values of FS's error probability and CE's NMSE are obtained by ``Prop'' for each given $K$ and compared with ``Ref\_\hspace{0.01em}\cite{ref_ML}'' and ``Ref\_\hspace{0.01em}\cite{ref_OMP}''. Thus, against the varying values of $K$, the proposed cascaded ELM-based JFSCE shows good robustness.
\begin{figure}[h]
\centering    %居中
\subfigure[Error probability of FS] %第一张子图
{
	\begin{minipage}{0.48\textwidth}
	\centering          %子图居中
	\includegraphics[scale=0.8]{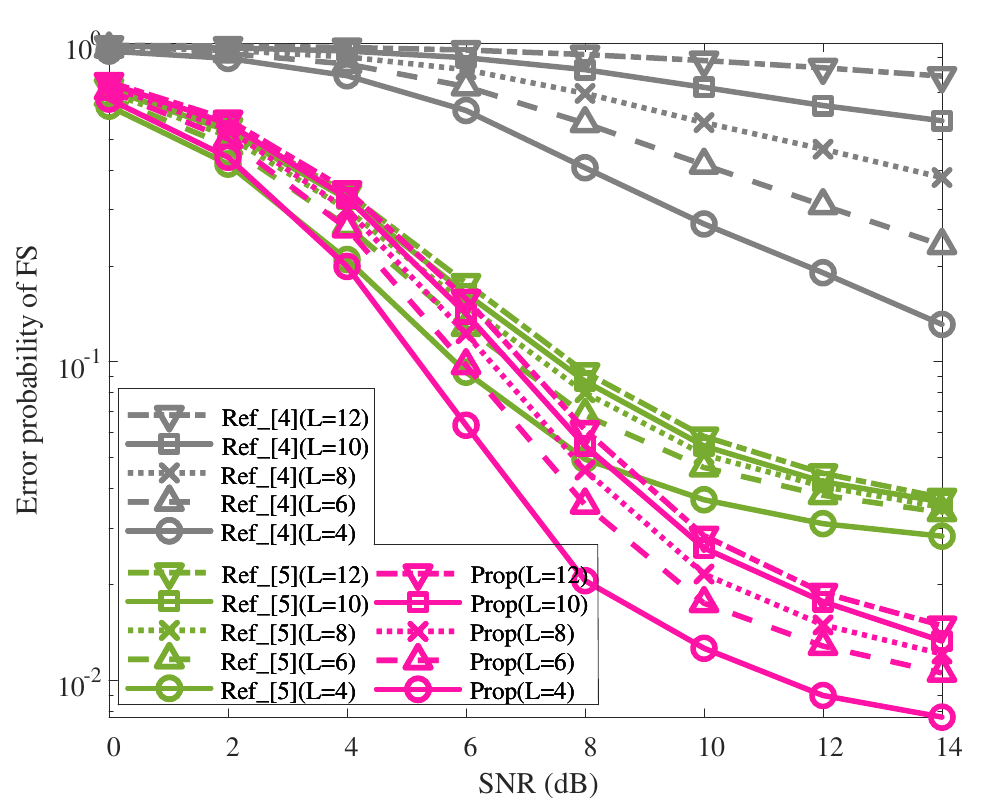}   %以pic.jpg的0.5倍大小输出
	\end{minipage}
	%\label{JFSCE_Vs_KKFS}
}	
\subfigure[NMSE of CE] %第二张子图
{
	\begin{minipage}{0.48\textwidth}
	\centering      %子图居中
	\includegraphics[scale=0.8]{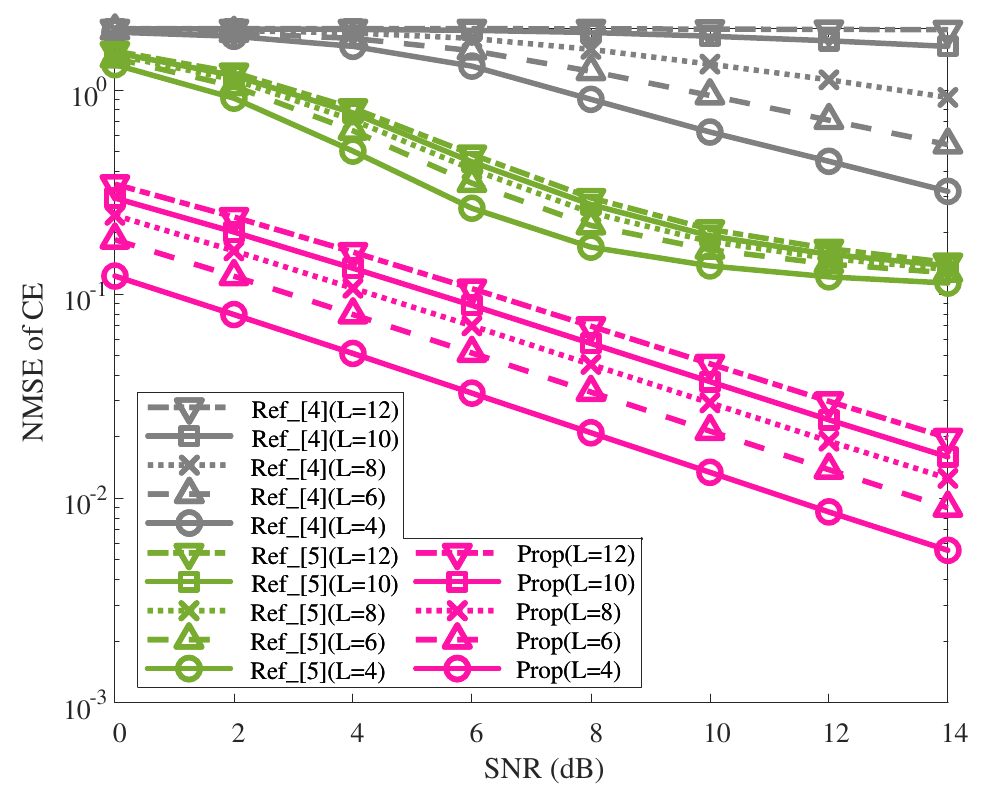}   %以pic.jpg的0.5倍大小输出
	\end{minipage}
	%\label{JFSCE_Vs_KKCE}  %图片引用标记
}
\caption{Error probability of FS and NMSE of CE against the impact of $L$, where $L=4$, $L=6$, $L=8$, $L=10$, and $L=12$.} %  %大图名称
\label{JFSCE_Vs_LLFSandCE}
\end{figure}
\subsubsection{Robustness against $L$}
For exhibiting the impact of different values of $L$, $L=4,L=6,L=8,L=10$, and $L=12$ are separately considered in Fig. \ref{JFSCE_Vs_LLFSandCE}. The FS's error probability and CE's NMSE are plotted in Fig. \ref{JFSCE_Vs_LLFSandCE}(a) and Fig. \ref{JFSCE_Vs_LLFSandCE}(b), respectively. From Fig. \ref{JFSCE_Vs_LLFSandCE}(a), the error probability of FS becomes worse with the increasing value of $L$ due to the increased multi-path interference. To make the ``Prop'' as an example, when ${\rm{SNR}}$ = 10dB and the values of $L$ are enlarged from 4 to 12 with the interval of 2, the values of FS's error probability are about $1.26 \times {10^{{\rm{ - }}2}}$, $1.75 \times {10^{{\rm{ - }}2}}$, $2.16 \times {10^{{\rm{ - }}2}}$, $2.61 \times {10^{{\rm{ - }}2}}$, and $2.85 \times {10^{{\rm{ - }}2}}$, respectively. Despite the value of FS's error probability in each given scheme (i.e., ``Prop'', ``Ref\_\hspace{0.01em}\cite{ref_ML}'', and ``Ref\_\hspace{0.01em}\cite{ref_OMP}'') rising with the increase of $L$, the ``Prop'' still achieves the minimal error probability for all given SNRs. For example, when $L= 8$ and ${\rm{SNR}}$ = 10dB, the ``Prop'' attains ${e_{{\rm{error}}}}$ = $2.16 \times {10^{{\rm{ - }}2}}$, yet the error probability of ``Ref\_\hspace{0.01em}\cite{ref_ML}'' and ``Ref\_\hspace{0.01em}\cite{ref_OMP}'' are about $5.64 \times {10^{{\rm{ - }}1}}$ and $5.11 \times {10^{{\rm{ - }}2}}$, respectively. From Fig. \ref{JFSCE_Vs_LLFSandCE}(b), for each given scheme, the NMSE of CE increases with the increase of $L$. For example, when ${\rm{SNR}}$ = 12dB, the NMSEs of ``Ref\_\hspace{0.01em}\cite{ref_ML}'' are about $4.47 \times {10^{{\rm{ - }}1}}$, $7.11 \times {10^{{\rm{ - }}1}}$, $1.12$, $1.74$, $1.98$ for the cases where $L = 4, 6, 8, 10, 12$. It could be also observed that, against the changing $L$, the minimal value of NMSE is still obtained by ``Prop'', and thus shows the best NMSE performance. For $L = 10$ and ${\rm{SNR}}$ = 10dB, the NMSE of the ``Prop'' reaches ${\overline \varepsilon _{{\rm{NMSE}}}}$ = $3.73 \times {10^{{\rm{ - }}2}}$, while ``Ref\_\hspace{0.01em}\cite{ref_ML}'' and ``Ref\_\hspace{0.01em}\cite{ref_OMP}'' cannot decrease to ${\overline \varepsilon _{{\rm{NMSE}}}}$ = $1 \times {10^{{\rm{ - }}1}}$. This reflects that, for different values of $L$, the ``Prop'' still achieves the optimum NMSE compared to ``Ref\_\hspace{0.01em}\cite{ref_ML}'' and ``Ref\_\hspace{0.01em}\cite{ref_OMP}''. In a word, compared with ``Ref\_\hspace{0.01em}\cite{ref_ML}'' and ``Ref\_\hspace{0.01em}\cite{ref_OMP}'', the FS's error probability and CE's NMSE are improved by the ``Prop'' against the impact of $L$. Therefore, with the change of $L$, the proposed cascaded ELM-based JFSCE shows good robustness.

\begin{figure}[h]
\centering    %居中
\subfigure[Error probability of FS] %第一张子图
{
	\begin{minipage}{0.48\textwidth}
	\centering          %子图居中
	\includegraphics[scale=0.8]{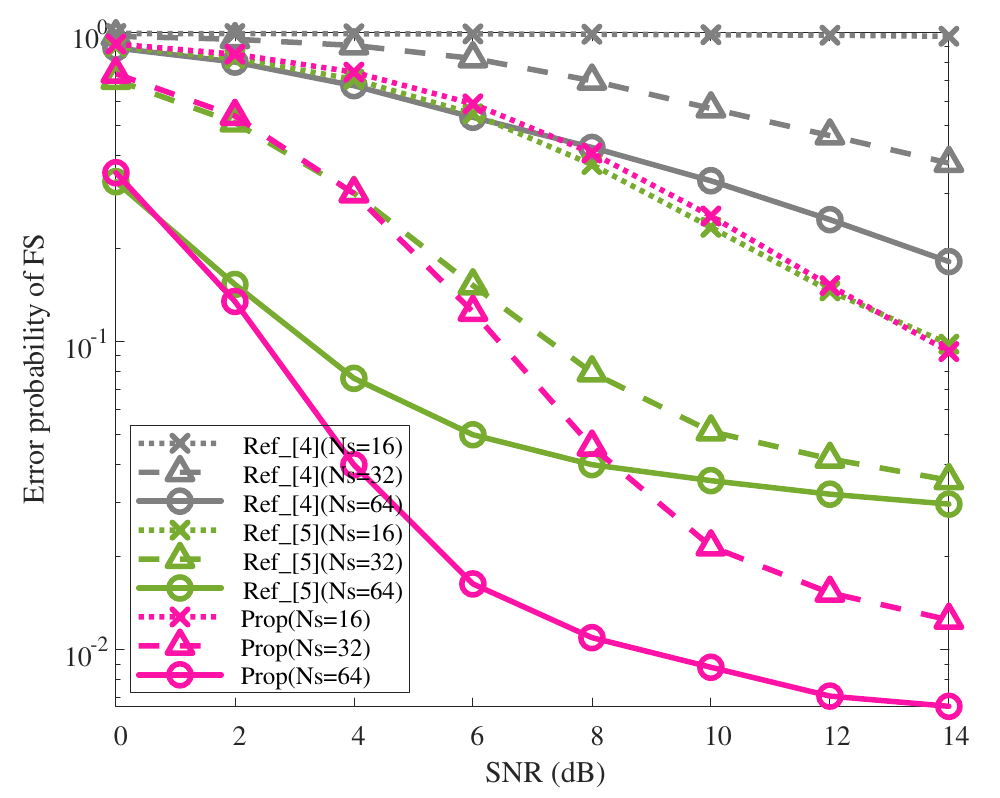}   %以pic.jpg 的0.5 倍大小输出
	\end{minipage}
	%\label{JFSCE_Vs_KKFS}
}	
\subfigure[NMSE of CE] %第二张子图
{
	\begin{minipage}{0.48\textwidth}
	\centering      %子图居中
	\includegraphics[scale=0.8]{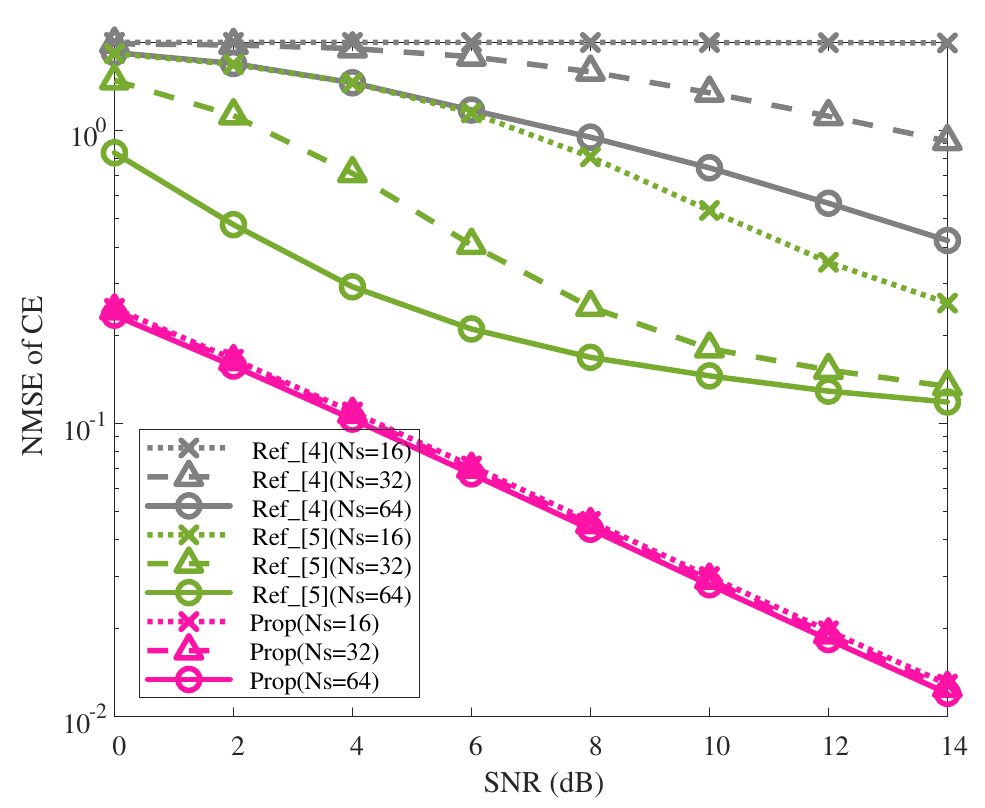}   %以pic.jpg 的0.5 倍大小输出
	\end{minipage}
	%\label{JFSCE_Vs_KKCE}  %图片引用标记
}
\caption{Error probability of FS and NMSE of CE against the impact of $N_s$, where $N_s=16$, $N_s=32$, and $N_s=64$.} %  %大图名称
\label{JFSCE_Vs_NsNsFSandCE}
\end{figure}

\subsubsection{Robustness against $N_s$}
In Fig. \ref{JFSCE_Vs_NsNsFSandCE}, to validate the robustness against $N_s$, FS's error probability and CE's NMSE are plotted, where $N_s = 16, N_s = 32$, and $N_s = 64$ are considered. From Fig. \ref{JFSCE_Vs_NsNsFSandCE}(a), with the increase of $N_s$, the FS's error probability declines for each given scheme due to the increased sufficiency of the training sequence. By using ``Ref\_\hspace{0.01em}\cite{ref_OMP}'' as an example, when $N_s = 32$ and ${\rm{SNR}}$ = 10dB, the values of FS's error probability are about $2.34 \times {10^{{\rm{ - }}1}}$, $5.10 \times {10^{{\rm{ - }}2}}$, and $3.53 \times {10^{{\rm{ - }}2}}$ for the values of $N_s$ are 16, 32, and 64. For the relatively small value of $N_s$, e.g., $N_s = 16$, the ``Prop'' maintains a comparable error probability of FS as that of ``Ref\_\hspace{0.01em}\cite{ref_OMP}'', while achieving a much smaller FS's error probability than that of ``Ref\_\hspace{0.01em}\cite{ref_ML}''. With the increase of $N_s$, the ``Prop'' obtains minimal error probability for the relatively large $N_s$, e.g., $N_s = $32, 64. For example, as $N_s = 32$ and ${\rm{SNR}}$ = 8dB, the error probability of ``Prop'' is about $4.55 \times {10^{{\rm{ - }}2}}$, yet the error probabilities of ``Ref\_\hspace{0.01em}\cite{ref_ML}'' and ``Ref\_\hspace{0.01em}\cite{ref_OMP}'' are about $6.98 \times {10^{{\rm{ - }}1}}$ and $7.92 \times {10^{{\rm{ - }}2}}$, respectively. From Fig. \ref{JFSCE_Vs_NsNsFSandCE}(b), as $N_s$ increases, the values of CE's NMSE for the given schemes decrease. For example, when ${\rm{SNR}}$ = 6dB for ``Ref\_\hspace{0.01em}\cite{ref_OMP}'', we have ${\overline \varepsilon _{{\rm{NMSE}}}}$ = $1.15$, $4.07 \times {10^{{\rm{ - }}1}}$, $2.10 \times {10^{{\rm{ - }}1}}$ for $N_s = 16, 32, 64$. Among these given schemes (i.e., ``Prop'', `Ref\_\hspace{0.01em}\cite{ref_ML}'' and ``Ref\_\hspace{0.01em}\cite{ref_OMP}''), the ``Prop'' still obtains the minimum of NMSE. For the case where $N_s = 32$ and ${\rm{SNR}}$ = 10dB, the NMSE of ``Prop'' is about $2.92 \times {10^{{\rm{ - }}2}}$, while the values of NMSE are about $1.35$ and $1.80 \times {10^{{\rm{ - }}1}}$ for ``Ref\_\hspace{0.01em}\cite{ref_ML}'' and ``Ref\_\hspace{0.01em}\cite{ref_OMP}'', respectively. This demonstrates the ``Prop'' owns a better NMSE than those of ``Ref\_\hspace{0.01em}\cite{ref_ML}'' and ``Ref\_\hspace{0.01em}\cite{ref_OMP}''. From Fig. \ref{JFSCE_Vs_NsNsFSandCE}(a) and Fig. \ref{JFSCE_Vs_NsNsFSandCE}(b), compared with ``Ref\_\hspace{0.01em}\cite{ref_ML}'' and ``Ref\_\hspace{0.01em}\cite{ref_OMP}'', the proposed cascaded ELM-based JFSCE achieves the best FS's error probability and CE's NMSE when facing the impact of $N_s$.

\begin{figure}[h]
\centering    %居中
\subfigure[Error probability of FS] %第一张子图
{
	\begin{minipage}{0.48\textwidth}
	\centering          %子图居中
	\includegraphics[scale=0.8]{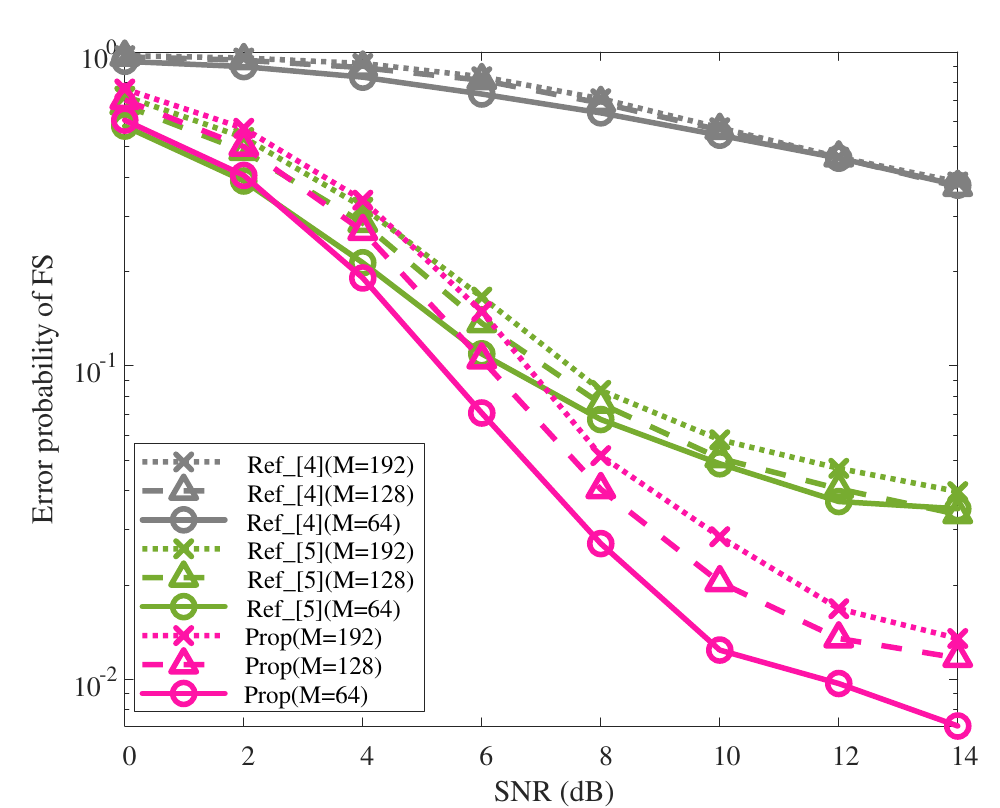}   %以pic.jpg的0.5倍大小输出
	\end{minipage}
	%\label{JFSCE_Vs_KKFS}
}	
\subfigure[NMSE of CE] %第二张子图
{
	\begin{minipage}{0.48\textwidth}
	\centering      %子图居中
	\includegraphics[scale=0.8]{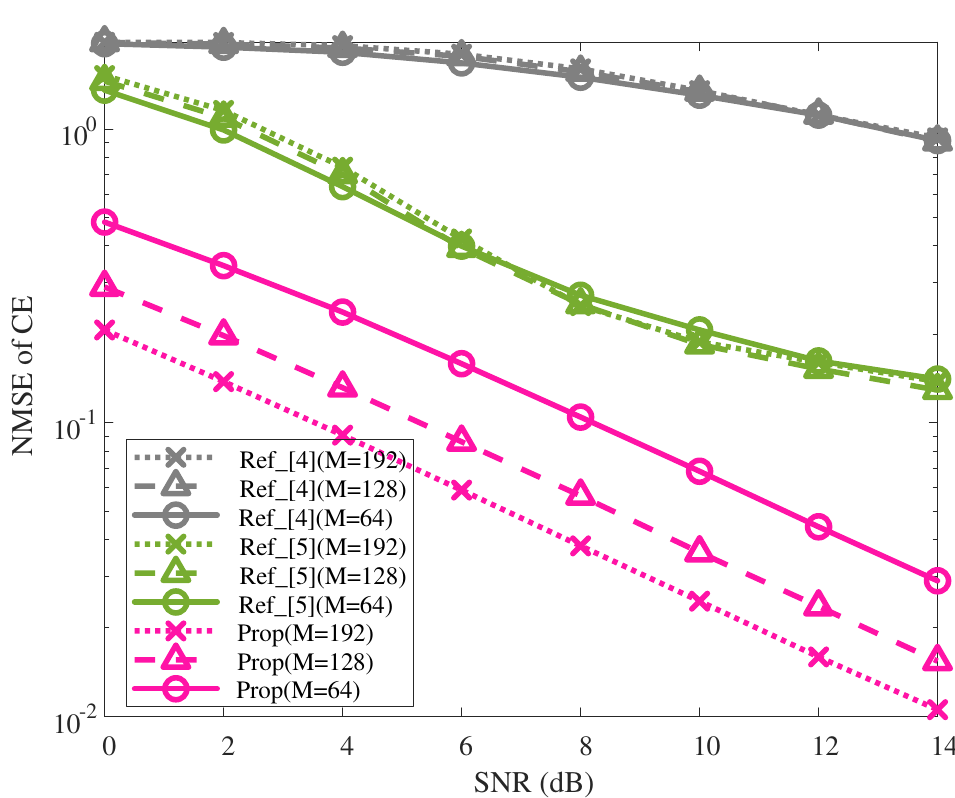}   %以pic.jpg的0.5倍大小输出
	\end{minipage}
	%\label{JFSCE_Vs_KKCE}  %图片引用标记
}
\caption{Error probability of FS and NMSE of CE against the impact of $M$, where $M=64$, $M=128$, and $M=192$.} %  % 大图名称
\label{JFSCE_Vs_NNFSandCE}
\end{figure}

\subsubsection{Robustness against $M$}
To demonstrate the impact of $M$, Fig. \ref{JFSCE_Vs_NNFSandCE} shows the error probability of FS and NMSE of CE, where $M = 64, M = 128$, and $M = 192$ are considered. From Fig. \ref{JFSCE_Vs_NNFSandCE}(a), with the increase of $M$, the FS's error probability increases due to the increased search space of FS. For example, when ${\rm{SNR}}$ = 8dB, the ``Ref\_\hspace{0.01em}\cite{ref_OMP}'' owns ${e_{{\rm{error}}}}$ = $6.74 \times {10^{{\rm{ - }}2}}$, $7.57 \times {10^{{\rm{ - }}2}}$, $8.35 \times {10^{{\rm{ - }}2}}$ for $M = 64, 128, 192$, respectively. In the relatively low SNR region (i.e., from 0dB to 6dB in Fig. \ref{JFSCE_Vs_NNFSandCE}(a)), similar error probabilities of FS are obtained by ``Prop'' and ``Ref\_\hspace{0.01em}\cite{ref_OMP}'', while it is much smaller than that of ``Ref\_\hspace{0.01em}\cite{ref_ML}''. In the region that ${\rm{SNR}}$ $\geq8$dB, with the increase of $M$, the ``Prop'' reaps the minimum of FS's error probability among the given FS schemes. For example, for the case where $M = 128$ and ${\rm{SNR}}$ = 10dB, the ``Prop'' achieves ${e_{{\rm{error}}}}$ = $2.04 \times {10^{{\rm{ - }}2}}$, while ``Ref\_\hspace{0.01em}\cite{ref_ML}'' and ``Ref\_\hspace{0.01em}\cite{ref_OMP}'' attain the error probability about $5.65 \times {10^{{\rm{ - }}1}}$ and $5.07 \times {10^{{\rm{ - }}2}}$, respectively. This reflects that, with the varying of $M$, then ``Prop'' achieves a lower FS's error probability than those of ``Ref\_\hspace{0.01em}\cite{ref_ML}'' and ``Ref\_\hspace{0.01em}\cite{ref_OMP}''. From Fig. \ref{JFSCE_Vs_NNFSandCE}(b), with the changing of $M$, the ``Prop'' achieves a significantly smaller NMSE than the other two given schemes in the whole given SNR region. For example, for $M = 128$ and ${\rm{SNR}}$ = 10dB, the NMSE of ``Prop'' is about $3.61 \times {10^{{\rm{ - }}2}}$, yet the NMSEs of ``Ref\_\hspace{0.01em}\cite{ref_ML}'' and ``Ref\_\hspace{0.01em}\cite{ref_OMP}'' are about $1.34$ and $1.84 \times {10^{{\rm{ - }}1}}$, respectively. This shows that, for different values of $M$, the ``Prop'' obtains the optimum NMSE among the three given schemes. From Fig. \ref{JFSCE_Vs_NNFSandCE}(a) and Fig. \ref{JFSCE_Vs_NNFSandCE}(b), compared with ``Ref\_\hspace{0.01em}\cite{ref_ML}'' and ``Ref\_\hspace{0.01em}\cite{ref_OMP}'', the proposed cascaded ELM-based JFSCE shows better performance of FS and CE against the variation of $M$.

%--------------------第五章 结论-------------------------------------------
\section{CONCLUSION}
In this paper, we propose the cascaded ELM-based JFSCE scheme in the scenarios of the Rician fading channel and HI. In this scheme, the initial features of FS and CE are extracted by using the conventional non-NN-based JFSCE method and then analyzed by the constructed FS-NET and CE-NET. Compared with the conventional JFSCE methods, the proposed cascaded ELM-based JFSCE can significantly reduce the error probability of FS and the NMSE of CE in the presence of HI. Against the impacts of varying parameters, the proposed scheme presents its effectiveness and robustness in LOS and NLOS scenarios. In future works, the ML-based JFSCE in orthogonal frequency division multiplex (OFDM) systems with nonlinear distortion will be investigated.

%--------------插入文献
%\nocite{*}
\bibliographystyle{ieeetran}
\bibliography{citeArticlesList}

\begin{thebibliography}{10}
\providecommand{\url}[1]{#1}
\csname url@samestyle\endcsname
\providecommand{\newblock}{\relax}
\providecommand{\bibinfo}[2]{#2}
\providecommand{\BIBentrySTDinterwordspacing}{\spaceskip=0pt\relax}
\providecommand{\BIBentryALTinterwordstretchfactor}{4}
\providecommand{\BIBentryALTinterwordspacing}{\spaceskip=\fontdimen2\font plus
\BIBentryALTinterwordstretchfactor\fontdimen3\font minus
  \fontdimen4\font\relax}
\providecommand{\BIBforeignlanguage}[2]{{%
\expandafter\ifx\csname l@#1\endcsname\relax
\typeout{** WARNING: IEEEtran.bst: No hyphenation pattern has been}%
\typeout{** loaded for the language `#1'. Using the pattern for}%
\typeout{** the default language instead.}%
\else
\language=\csname l@#1\endcsname
\fi
#2}}
\providecommand{\BIBdecl}{\relax}
\BIBdecl

\bibitem{FSCE1}
M.~Speth, F.~Classen, and H.~Meyr, ``Frame synchronization of ofdm systems in
  frequency selective fading channels,'' in \emph{IEEE 47 VTC}, vol.~3, Jun.
  1997, pp. 1807--1811 vol.3.

\bibitem{FSCE2}
H.~Ji, S.~Park, J.~Yeo, Y.~Kim, J.~Lee, and B.~Shim, ``Ultra-reliable and
  low-latency communications in 5g downlink: Physical layer aspects,''
  \emph{IEEE Wireless Commun.}, vol.~25, no.~3, pp. 124--130, Jun. 2018.

\bibitem{flat_fading}
Q.~Zhao, Z.~Zhou, J.~Li, and B.~Vucetic, ``Joint semi-blind channel estimation
  and synchronization in two-way relay networks,'' \emph{IEEE Trans. Veh.
  Technol.}, vol.~63, no.~7, pp. 3276--3293, Sep. 2014.

\bibitem{ref_ML}
Y.~Wang, K.~Shi, and E.~Serpedin, ``Continuous-mode frame synchronization for
  frequency-selective channels,'' \emph{IEEE Trans. Veh. Technol.}, vol.~53,
  no.~3, pp. 865--871, May. 2004.

\bibitem{ref_OMP}
z.~Özdemir, R.~Hamila, N.~Al-Dhahir, and I.~Güvenç, ``Sparsity-aware joint
  frame synchronization and channel estimation: Algorithm and usrp
  implementation,'' in \emph{Proc. 2017 IEEE Military Communications Conference
  (MILCOM 2017)}, Oct. 2017, pp. 647--652.

\bibitem{HI_r1}
L.~Tlebaldiyeva, B.~Maham, and T.~A. Tsiftsis, ``Device-to-device mmwave
  communication in the presence of interference and hardware distortion
  noises,'' \emph{IEEE Commun. Lett.}, vol.~23, no.~9, pp. 1607--1610, Sep.
  2019.

\bibitem{HI_r2}
K.~Guo, K.~An, F.~Zhou, T.~A. Tsiftsis, G.~Zheng, and S.~Chatzinotas, ``On the
  secrecy performance of noma-based integrated satellite multiple-terrestrial
  relay networks with hardware impairments,'' \emph{IEEE Trans. Veh. Technol.},
  vol.~70, no.~4, pp. 3661--3676, 2021.

\bibitem{low_cost}
S.~V. Kulygin and V.~O. Kazachkov, ``Modeling of nonlinear distortions in 5g nr
  systems,'' in \emph{Proc. SSGPFBC}, Mar. 2021, pp. 1--4.

\bibitem{energy}
M.~B. Salman and G.~M. Guvensen, ``An efficient qam detector via nonlinear
  post-distortion based on fde bank under pa impairments,'' \emph{IEEE Trans.
  Commun.}, pp. 1--1, Jul. 2021.

\bibitem{compute}
V.~V. Kirillov and P.~A. Turalchuk, ``Analysis of nonlinear distortions in
  transmitarrays,'' in \emph{Proc. ElConRus}, Jan. 2021, pp. 133--136.

\bibitem{c9}
C.~An, B.~Kim, and H.-G. Ryu, ``Design of w-ofdm and nonlinear performance
  comparison for 5g waveform,'' in \emph{Proc. ICTC}, Oct. 2016, pp.
  1006--1009.

\bibitem{c10}
O.~B.~H. Belkacem, M.~L. Ammari, and R.~Dinis, ``Performance analysis of noma
  in 5g systems with hpa nonlinearities,'' \emph{IEEE Access}, vol.~8, pp.
  158\,327--158\,334, Aug. 2020.

\bibitem{c11}
R.~Guo, K.~Wang, Z.~Deng, W.~Lin, and R.~Song, ``A prediction model for channel
  state information in satellite communication system,'' in \emph{Proc.
  fPIMRC}, Oct. 2020, pp. 1--6.

\bibitem{MLtoHI2}
C.~Qing, L.~Dong, L.~Wang, J.~Wang, and C.~Huang, ``Joint model and data driven
  receiver design for data-dependent superimposed training scheme with
  imperfect hardware,'' \emph{IEEE Trans. Wirel. Commun.}, pp. 1--1, Nov. 2021.

\bibitem{MLtoHI3}
B.~{Lim}, W.~J. {Yun}, J.~{Kim}, and Y.-C. {Ko}, ``{Joint Pilot Design and
  Channel Estimation using Deep Residual Learning for Multi-Cell Massive MIMO
  under Hardware Impairments},'' \emph{arXiv e-prints}, p. arXiv:2108.04485,
  Aug. 2021.

\bibitem{arXiv200709248L}
J.~{Liu}, K.~{Mei}, X.~{Zhang}, D.~{McLernon}, D.~{Ma}, J.~{Wei}, and S.~A.~R.
  {Zaidi}, ``{Fine Timing and Frequency Synchronization for MIMO-OFDM: An
  Extreme Learning Approach},'' \emph{arXiv e-prints}, p. arXiv:2007.09248,
  Jul. 2020.

\bibitem{OFDM_time_synchron}
C.~{Qing}, S.~{Tang}, C.~{Rao}, Q.~{Ye}, J.~{Wang}, and C.~{Huang}, ``{Label
  Design-based ELM Network for Timing Synchronization in OFDM Systems with
  Nonlinear Distortion},'' \emph{arXiv e-prints}, p. arXiv:2107.13177, Jul.
  2021.

\bibitem{ELM_fs1}
C.~Qing, W.~Yu, B.~Cai, J.~Wang, and C.~Huang, ``Elm-based frame
  synchronization in burst-mode communication systems with nonlinear
  distortion,'' \emph{IEEE Wireless Commun. Lett.}, vol.~9, no.~6, pp.
  915--919, Jun. 2020.

\bibitem{ELM_fs2}
C.~Qing, W.~Yu, S.~Tang, C.~Rao, and J.~Wang, ``Elm-based frame synchronization
  in nonlinear distortion scenario using superimposed training,'' \emph{IEEE
  Access}, vol.~9, pp. 53\,530--53\,539, Apr. 2021.

\bibitem{YW_ref1}
F.~Ling, \emph{Synchronization in digital communication systems}.\hskip 1em
  plus 0.5em minus 0.4em\relax Cambridge University Press, Jun. 2017.

\bibitem{YW_ref2}
B.~Lopes, S.~Catarino, N.~M.~B. Souto, R.~Dinis, and F.~Cercas, ``Robust joint
  synchronization and channel estimation approach for frequency-selective
  environments,'' \emph{IEEE Access}, vol.~6, pp. 53\,180--53\,190, Sep. 2018.

\bibitem{huang2004extreme}
G.-B. Huang, Q.-Y. Zhu, and C.-K. Siew, ``Extreme learning machine: a new
  learning scheme of feedforward neural networks,'' in \emph{2004 IEEE
  international joint conference on neural networks (IEEE Cat. No. 04CH37541)},
  vol.~2.\hskip 1em plus 0.5em minus 0.4em\relax Ieee, Jul. 2004, pp. 985--990.

\bibitem{huang2006extreme}
G.-B. \vspace{0mm} Huang, Q.-Y. Zhu, and C.-K. Siew, ``Extreme learning
  machine: theory and applications,'' \emph{Neurocomputing}, vol.~70, no. 1-3,
  pp. 489--501, 2006.

\bibitem{indoors1}
K.-W. Yip, Y.-C. Wu, and T.-S. Ng, ``Timing-synchronization analysis for ieee
  802.11a wireless lans in frequency-nonselective rician fading environments,''
  \emph{IEEE Trans. Wireless Commun.}, vol.~3, no.~2, pp. 387--394, Mar. 2004.

\bibitem{distance_shorter}
Y.~Zhang, L.~Yang, and H.~Zhu, ``Cell-free massive mimo systems with
  low-resolution adcs: The rician fading case,'' \emph{IEEE Syst J}, pp. 1--12,
  Jan.2021.

\bibitem{A2G}
Y.~Liu, K.~Xiong, Y.~Lu, Q.~Ni, P.~Fan, and K.~B. Letaief, ``Uav-aided wireless
  power transfer and data collection in rician fading,'' \emph{IEEE J. Sel.
  Areas Commun.}, pp. 1--1, Jul. 2021.

\bibitem{millimeterWave}
M.~K. Samimi, G.~R. MacCartney, S.~Sun, and T.~S. Rappaport, ``28 ghz
  millimeter-wave ultrawideband small-scale fading models in wireless
  channels,'' in \emph{Proc. VTC Spring}, May. 2016, pp. 1--6.

\bibitem{rice_discrip}
L.~Sun, J.~Hou, and T.~Shu, ``Bandwidth-efficient precoding in cell-free
  massive mimo networks with rician fading channels,'' in \emph{Proc. SECON},
  Jul. 2021, pp. 1--9.

\bibitem{act_function}
T.~V. Luong, Y.~Ko, N.~A. Vien, D.~H.~N. Nguyen, and M.~Matthaiou, ``Deep
  learning-based detector for ofdm-im,'' \emph{IEEE Wireless Commun. Lett.},
  vol.~8, no.~4, pp. 1159--1162, Aug. 2019.

\bibitem{rician_fading}
P.~Liu, D.~Kong, J.~Ding, Y.~Zhang, K.~Wang, and J.~Choi, ``Channel estimation
  aware performance analysis for massive mimo with rician fading,'' \emph{IEEE
  Trans. Commun.}, vol.~69, no.~7, pp. 4373--4386, Jul. 2021.

\bibitem{sigmoid_fun}
B.~Ding, H.~Qian, and J.~Zhou, ``Activation functions and their characteristics
  in deep neural networks,'' in \emph{2018 Chinese Control And Decision
  Conference (CCDC)}, Jun. 2018, pp. 1836--1841.

\bibitem{DeepLearning2NonlinearDis2}
C.~Qing, B.~Cai, Q.~Yang, J.~Wang, and C.~Huang, ``Deep learning for csi
  feedback based on superimposed coding,'' \emph{IEEE Access}, vol.~7, pp.
  93\,723--93\,733, Jul. 2019.

\bibitem{c20}
D.~Chu, ``Polyphase codes with good periodic correlation properties
  (corresp.),'' \emph{IEEE Trans. Inf. Theory}, vol.~18, no.~4, pp. 531--532,
  Jul. 1972.

\bibitem{EVM_formula}
A.~M. Angelotti, G.~P. Gibiino, C.~Florian, and A.~Santarelli, ``Broadband
  error vector magnitude characterization of a gan power amplifier using a
  vector network analyzer,'' in \emph{2020 IEEE/MTT-S International Microwave
  Symposium (IMS)}, 2020, pp. 747--750.

\bibitem{TWT_params}
A.~Saleh, ``Frequency-independent and frequency-dependent nonlinear models of
  twt amplifiers,'' \emph{IEEE Trans.Commun.}, vol.~29, no.~11, pp. 1715--1720,
  1981.

\bibitem{EVM_saturation}
H.~Abdulkader, F.~Langlet, D.~Roviras, and F.~Castani{\'e}, ``Natural gradient
  algorithm for neural networks applied to non-linear high power amplifiers,''
  \emph{International Journal of Adaptive Control and Signal Processing},
  vol.~16, no.~8, pp. 557--576, Sep. 2002.

\bibitem{ref_DNN}
A.~Melgar, A.~de~la Fuente, L.~Carro-Calvo, {\'O}.~Barquero-P{\'e}rez, and
  E.~Morgado, ``Deep neural network: an alternative to traditional channel
  estimators in massive mimo systems,'' \emph{IEEE Transactions on Cognitive
  Communications and Networking}, vol.~8, no.~2, pp. 657--671, Jun. 2022.

\bibitem{multiply_complex}
Z.~Zhang, C.~Gong, Y.~Dong, X.~Wang, and X.~Dai, ``Expectation propagation
  aided signal detection for uplink massive generalized spatial modulation mimo
  systems,'' \emph{IEEE Trans. Wireless Commun.}, vol.~21, no.~3, pp.
  2006--2018, Sep. 2021.

\bibitem{FC_complex}
J.~Guo, C.-K. Wen, and S.~Jin, ``Canet: Uplink-aided downlink channel
  acquisition in fdd massive mimo using deep learning,'' \emph{IEEE
  Transactions on Communications}, vol.~70, no.~1, pp. 199--214, 2021.

\end{thebibliography}

\begin{IEEEbiography}[{\includegraphics[width=1in,height=1.25in,clip,keepaspectratio]{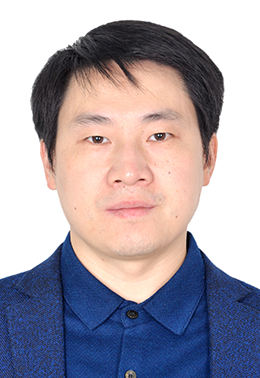}}]{Chaojin Qing}
(M'15) received the B.S. degree in communication engineering from Chengdu University of Information Technology, Chengdu, China, in 2001, the M.S. and Ph.D. degrees in communications and information systems from the University of Electronic Science and Technology of China, Chengdu, China, in 2006 and 2011, respectively. From November 2015 to December 2016, he was a Visiting Scholar with Broadband Communication Research Group (BBCR) of the University of Waterloo, Waterloo, ON, Canada.

From 2001 to 2004, he was a teacher with the Communications Engineering Teaching and Research Office, Chengdu University of Information Technology, Chengdu, China. Since 2011, he has been a Professor with the School of Electrical Engineering and Electronic Information, Xihua University, Chengdu, China. He is the author of more than 50 papers and more than 20 chinese inventions. His research interests include detection and estimation, massive MIMO systems, and deep learning in physical layer of wireless communications.
\end{IEEEbiography}

\begin{IEEEbiography}[{\includegraphics[width=1in,height=1.25in,clip,keepaspectratio]{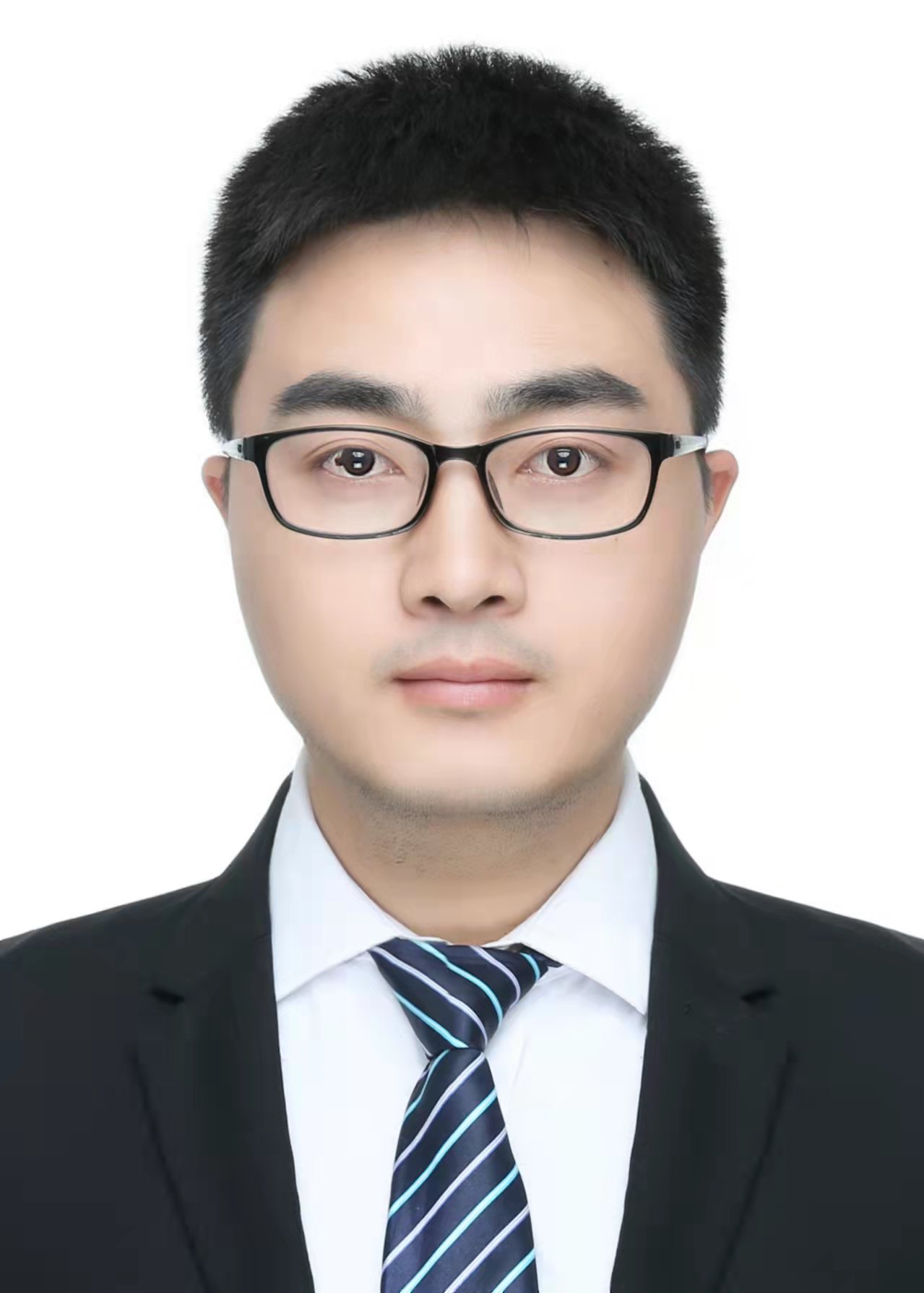}}]{Chuangui Rao}received the B. S. degree in electrical engineering and information from Southwest Petroleum University, China, in 2017. He is currently pursuing the M.S. degree with the School of Electrical Engineering and Electronic Information, Xihua University, Chengdu, China, under the supervision of Prof. Qing. His research interests include frame synchronization and channel estimation, and deep learning applications in physical layer of wireless communications.
\end{IEEEbiography}

\begin{IEEEbiography}[{\includegraphics[width=1in,height=1.25in,clip,keepaspectratio]{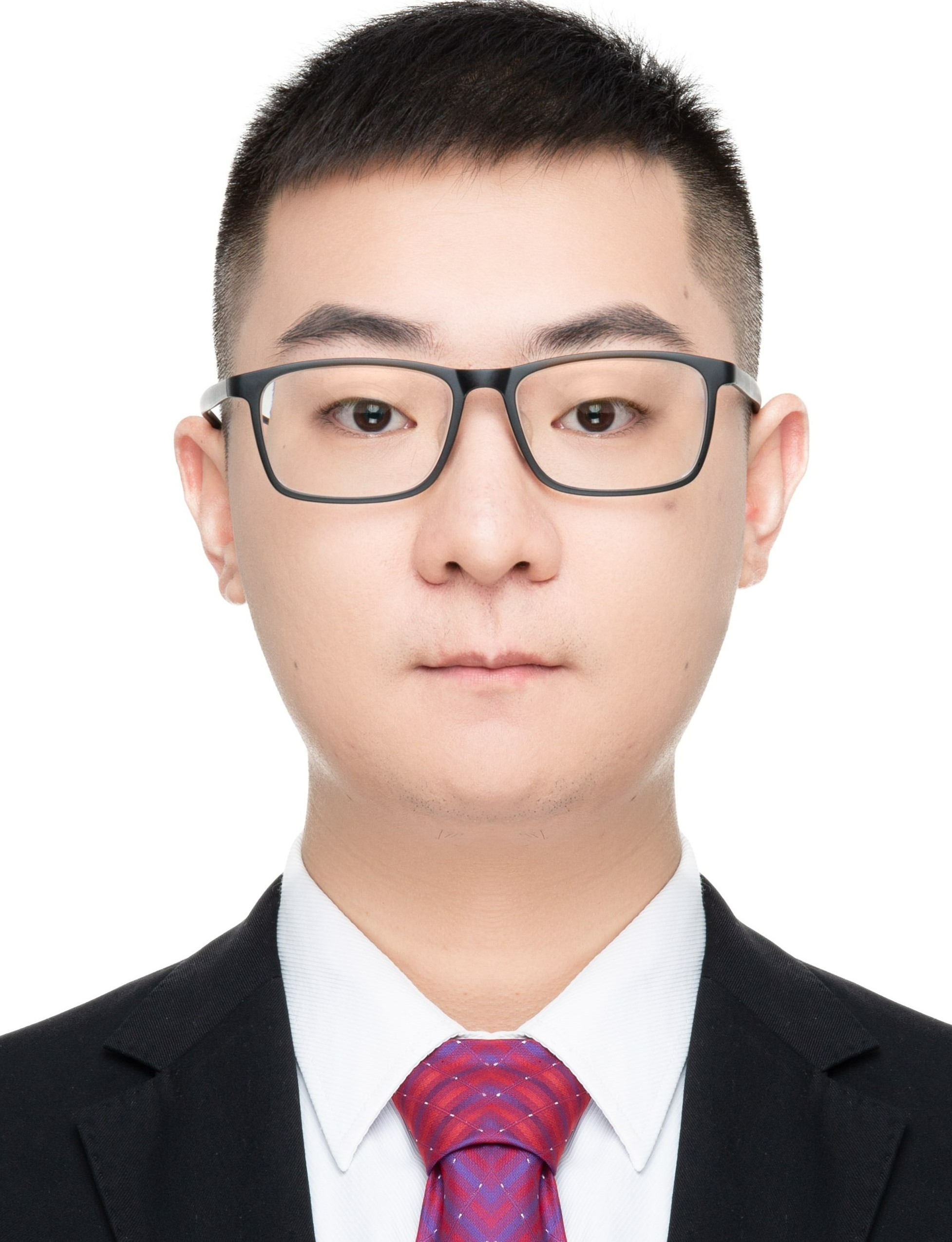}}]{ShuHai Tang}received the B. S. degree from the School of Electrical Engineering and Electrical Information, Xihua University, Chengdu, China, in 2020, where he is currently pursuing the M. S. degree under the supervision of Prof. Qing. His research interests include synchronization and channel estimation in OFDM system, and machine learning applications in physical layer of wireless communications.
\end{IEEEbiography}

\begin{IEEEbiography}[{\includegraphics[width=1in,height=1.25in,clip,keepaspectratio]{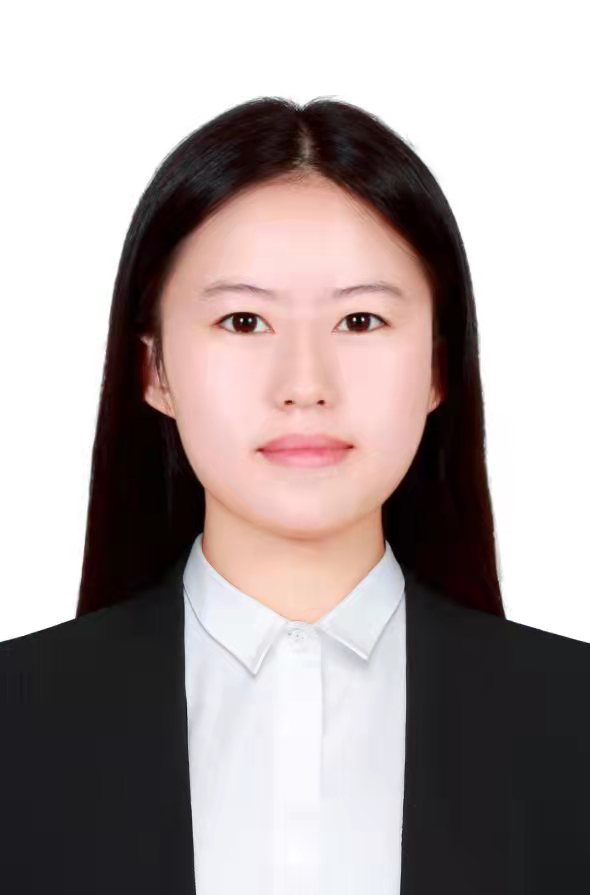}}]{Na Yang}received the B. S. degree from the School of Electrical Engineering and Electrical Information, Xihua University, Chengdu, China, in 2021, where she is currently pursuing the M. S. degree under the supervision of Prof. Qing. Her research interests include synchronization in OTFS and OTSM system, and machine learning applications in physical layer of wireless communications.
\end{IEEEbiography}

\begin{IEEEbiography}[{\includegraphics[width=1in,height=1.25in,clip,keepaspectratio]{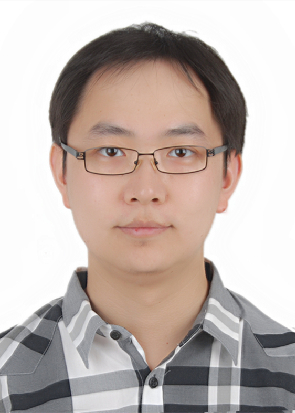}}]{Jiafan Wang} (S'15) received his B.S. degree and M.S. degree in Electrical Engineering from University of Electronic Science and Technology of China in 2006 and 2009, respectively. He accomplished the Ph.D.  degree in Computer Engineering at Texas A$\&$M University, College Station, TX, USA in 2017.

He is now working as an AI Software Development Engineer in a high-tech company and as the technical consultant of the Wireless Networking \& Communication Group of Xihua University. The major responsibility of his work is to develop and optimize machine learning models, including, but not limited to computer vision, image processing, natural language processing, deep learning recommendation, etc. The optimization is achieved by compiling the machine learning models into internal graph nodes and performing the computation upon high efficient distributed hardware. He will also analyze and maintain the software and hardware thus guaranteeing the acceleration of model training and inferencing.
\end{IEEEbiography}

\end{document}